\documentclass[aps,preprintnumbers,  amsmath, showpacs]{revtex4}
\usepackage{psfrag}
\usepackage{epsfig}
\usepackage{amsfonts} 
\usepackage{amssymb}  
\usepackage{graphicx} 
\usepackage{amsmath}  
\usepackage{amsmath}  
\usepackage{calligra}
\usepackage{mathrsfs}
\usepackage{tabularx}
\usepackage[latin1]{inputenc}
\usepackage{color}
\usepackage{suffix}
\usepackage{mathtools}
\usepackage{slashed}
\usepackage{ulem}

\DeclareMathAlphabet{\mathcalligra}{T1}{calligra}{m}{n}
\DeclareFontShape{T1}{calligra}{m}{n}{<->s*[2.2]callig15}{}

\newcommand{\bea}{\begin{eqnarray}}
\newcommand{\eea}{\end{eqnarray}}
\newcommand{\beas}{\begin{eqnarray*}}
\newcommand{\eeas}{\end{eqnarray*}}

\DeclarePairedDelimiterX\MeijerM[3]{\lparen}{\rparen}%
{\begin{smallmatrix}#1 \\ #2\end{smallmatrix}\delimsize\vert\,#3}
\DeclareMathOperator{\Tr}{Tr}
\newcommand\MeijerG[8][]{%
  G^{\,#2,#3}_{#4,#5}\MeijerM[#1]{#6}{#7}{#8}}

\WithSuffix\newcommand\MeijerG*[7]{%
  G^{\,#1,#2}_{#3,#4}\MeijerM*{#5}{#6}{#7}}

\def\Title#1{\begin{center} {\Large {\bf #1} } \end{center}}

\begin{document}

\Title{Hadron-Quark Phase Transition at Finite Density in the Presence of a Magnetic Field: Anisotropic Approach}

\author{E. J. Ferrer and A. Hackebill}
\affiliation{Dept. of Physics and Astronomy, University of Texas Rio Grande Valley, Edinburg 78539, USA, 
Physics Department, CUNY-Graduate Center, New York 10314, USA}
 
\begin{abstract}
We investigate the hadron-quark phase transition at finite density in the presence of a magnetic field taking into account the anisotropy created  by a uniform magnetic field in the system's equations of state. We find a new anisotropic equilibrium condition that will drive the first-order phase transition along the boundary between the two phases. Fixing the magnetic field in the hadronic phase, the phase transition is realized by increasing the baryonic chemical potential at zero temperature. It is shown that the magnetic field is mildly boosted after the system transitions from the hadronic to the quark phase. The magnetic-field discontinuity between the two phases is supported by a surface density of magnetic monopoles, which accumulate at the boundary separating the two phases. The mechanism responsible for the monopole charge density generation is discussed. Each phase is found to be paramagnetic with higher magnetic susceptibility in the quark phase. The connection with the physics of neutron stars is highlighted through out the paper.
\end{abstract}

\pacs{05.70.Ce, 05.70.Fh, 21.65.-f, 26.60.-c}


\maketitle

\section{Introduction}

Isolated quarks cannot be seen
in nature. The reason for this is that quarks forming color neutral bound states, called hadrons, are favored at low energies. This phenomenon is called color
confinement. However, at sufficiently high temperatures, which occur for example in heavy-ion collisions, the bound states will eventually
melt forming a state called quark-gluon plasma, where quarks and gluons become weakly coupled due to asymptotic freedom. 

On the other hand, it is also accepted that for strongly interacting matter a transition from a color insulator (hadronic
matter) to a color conductor (quark-gluon plasma), takes place once the density of color charges is
sufficiently high to make the Debye radius smaller than the  hadron radius \cite{Deconfinement}. Since the Debye radius decreases with matter density, at sufficiently high density,
the long-range confining part of the interacting color potential becomes screened and it  also yields to quark deconfinement.

The natural system where deconfinement may be realized due to high densities is the core of a neutron star (NS). The matter density in the interior of NS increases from its surface to its core reaching values that possibly exceed that of nuclear matter $\rho_{nuc}=2.8\times 10^{14}$ g/cm$^3$ \cite{Shapiro}. At such densities, individual nucleons overlap substantially implying that under such conditions matter might consist of deconfined quarks rather than hadrons \cite{Collins}.

Since baryons can only be created and annihilated in pair with anti-baryons, the baryon number $N_B$, which is the number of baryons minus
the number of anti-baryons, is a conserved quantum number that maintains its value even after deconfinement. If $N_B$ is allowed to vary in the system, we can introduce a baryonic chemical potential $\mu$ as a Lagrange multiplier in the
grand canonical ensemble. For large baryon number, $\mu$ will be large, and in principle the critical value where the phase transition from hadronic mater to quark matter takes place can be reached.

When considering the inner medium of compact stars as a natural medium to study quark deconfinement, we should take into account another ingredient that is also present in that context: a strong magnetic field. It has been found, from the measured  periods  and  spin  down  of  soft-gamma  repeaters  (SGR)  and  anomalous  X-ray pulsars (AXP), as well as from the observed X-ray luminosities of AXP, that a certain class of NS named magnetars can have surface magnetic fields as large as $10^{14}$-$10^{16}$G  \cite{Soft-gamma repeaters}. Moreover, since the stellar medium has a very high electric conductivity, the magnetic flux should be conserved. Hence, the magnetic field strength increases with increasing matter density, and consequently a much stronger magnetic field in the stars' inner core should be expected. Although, the inner magnetic field of NS is not accessible to observation, its value has been estimated. Estimates based on the equipartition of energy principle between gravity and nuclear \cite{Nuclear-B}, or quark matter and considering both gravitationally bound and self-bound stars \cite{magnetizedfermions}, have led to maximum fields $\sim10^{18}$ and $\sim 10^{20}$G, respectively. Nevertheless, when the hydrodynamic equilibrium between gravity and matter pressures has been taken into account following different approaches \cite{Magnetohydro-Eq}-\cite{neutrons},  the resultant maximum field value for stable configurations has been of order $eB \sim 10^{17}$ G. 

Using QCD inspired models, several studies have suggested that the transition from hadronic matter to quark matter at high baryon chemical potential  \cite{First-Trans}, as well as in the presence of a magnetic field at finite temperature and zero density \cite{Agasian}, is a first-order phase transition. The main goal of the present paper is to investigate how the magnetic field affects such a  transition. Attempts to answer this question already exist in the literature \cite{Trans-B, Rabhi}. But in all those previous works an isotropic Maxwell or Gibbs mechanical equilibrium condition  was used, where the pressure of each magnetized phase was the same in all directions. Nevertheless, it has been shown \cite{Canuto, magnetizedfermions} that for a system of charged fermions the pressure in the presence of a uniform magnetic field becomes anisotropic having different values along the direction and transverse to the direction of the magnetic field. Later on, in Ref. \cite{neutrons}, the pressure anisotropy was investigated for neutral composite fermions (neutrons), where the interaction with the magnetic field was through the particle anomalous magnetic moment (AMM).  In  Refs.  \cite{magnetizedfermions, neutrons},  the  transverse  and  longitudinal  pressures  were  found  by  taking  the quantum-statistical average of the energy-momentum tensor using the path-integral approach.  The obtained results coincide with those obtained years ago by using the many-particle density matrix in a second-quantization approach \cite{Canuto}.  Thus, the main goal of the present paper is to investigate the consequences of considering an anisotropic equilibrium condition in the hadron-quark phase transition. We will use in particular the Maxwell construction \cite{Glendenning} with local charge conservation.  A major result in this regard is that in passing from the hadronic phase to the quark phase the magnetic field will be boosted by an increment that depends on how paramagnetic the quark phase is relative to the hadronic phase. That is, the difference between the magnetization of the two phases determines the jump in the magnetic field. As a consequence of that we will discuss how at the boundary between the two phases a density of magnetic monopoles will be accumulated.
 
 The physical frame-work of the system under consideration is as follows: First, in the hadronic phase we will use the nonlinear Walecka (NLW)  model \cite{Walecka}, where the interaction between the hadrons is mediated by several meson fields. We will calculate the energy density and pressures for these new particles starting with their corresponding stress-energy tensor as was done in 
 \cite{magnetizedfermions, neutrons} for fermions. We will show, that in the mean-field approximation, there is no anisotropy in the meson pressures, which is in agreement with the fact that the considered meson fields are electrically neutral and do not interact with the applied magnetic field. Second, we will show that the condensates of the zero components of the meson fields enter as new components of the effective chemical potential. Third, we will consider, for the sake of simplicity, that the quark deconfinement will give rise to a free-quark system described by the MIT bag model. Nevertheless, the possibility of having other magnetized phases after the transition such as those with inhomogeneous chiral condensates \cite{Frolov}-\cite{MDCDW} or color superconductivity \cite{MCFL} are feasible. This will be a valuable goal for future investigations.
 
Finally, we derived the anisotropic thermodynamic pressures, which originate in the presence of a uniform magnetic field, using thermodynamic principles. On the same footing, we derive the anisotropic equilibrium conditions characterizing a magnetized thermodynamic system. We show that the thermodynamic pressures derived for an anisotropic magnetized system coincide with those obtained from the quantum-staistical average of the stress-energy tensor.

The paper is organized as follows: In Sec. \ref{section2}, for the sake of clarity and to facilitate the reader's understanding, we review the main ideas that yield an anisotropic equation of state (EOS) in the presence of a magnetic field. In particular, we will show how the pressures obtained from the quantum-statistical average of the energy-momentum tensor coincide with the thermodynamic pressures derived from the thermodynamic potentials. In Sec. III, we derive from the equilibrium thermodynamic equations the anisotropic equilibrium conditions valid for a magnetized system of charged relativistic particles. Using the obtained mechanical equilibrium conditions in the Maxwell construction we show that the magnetic field value jumps through the boundary of the first-order phase transition. We discuss how the difference between the two magnetic fields positioned on both sides of the border between the hadronic phase (HP) and the quark phase (QP) induce a density of magnetic monopoles in that region. In Sec. IV, we discuss how the EOS for a magnetized-dense-hadron system with meson interactions, as well as for a magnetized system of free quarks,  can be derived from quantum statistical averaging of the energy-momentum tensor and we explicitly calculate the EOS for the NLW and MIT bag models in a uniform background magnetic field. At the end of this section, numerical calculations are reported showing how the different thermodynamic parameters behave around the phase-transition boundary. In Sec. V, we determine the magnetic susceptibility of the HP and QP, and we discuss how it helps explain our results from Sec. IV. In Sec. VI, we make our concluding remarks and indicate some possible future directions for the investigations of hadron-quark phase transitions in strong magnetic fields. In Appendix A, we derive from first principles, the meson contribution to the system EOS in the mean-field approximation. Finally, in Appendix B we derive the main formulas we use for the thermodynamic potential in the weak-field approximation.

\section{Anisotropic equations of state of dense matter in a magnetic field} \label{section2}

In this section, we review for the sake of clarity and a better understanding the EOS's of a relativistic system of charged particles in a magnetic field as was introduced in Ref. \cite{magnetizedfermions}.
To this end, we consider that the EOS of a magnetized system can be obtained through the quantum-statistical averaging of the components of the symmetrized field theoretic energy momentum tensor.  From the components of the average stress-energy tensor, we can find the system pressures and energy density.

\subsection{Covariant structure of the energy-momentum tensor at B$\neq$ 0}

 In \cite{magnetizedfermions}, it was found that in a magnetized medium the energy-momentum tensor has three independent structures that in a covariant formulation are given by
 \begin{equation}\label{Tau}
\tau^{\mu \nu}=a_1 \eta^{\mu \nu}+a_2u^\mu u^\nu +a_3\widehat{F}^{\mu \rho}\widehat{F}_{\rho}^\nu,
\end{equation}
 where $\eta^{\mu\nu}$ is the Minkowski metric, $u_\mu$ is the medium four-velocity, which in the rest frame takes the value $u_\mu =(1,\overrightarrow{0})$,  $\widehat{F}^{\mu \rho}=F^{\mu \rho} /B$ is the normalized electromagnetic strength tensor and $a_i$, $i=1,2,3$ are scalar coefficients, which when defining the quantum-statistical average of the energy-momentum tensor, $\big<\hat{\tau}_{\mu\nu} \big>$, depend on the external parameters such as temperature, chemical potentials, magnetic field, etc.
 
The existence of the three independent structures in (\ref{Tau}) is in agreement with the symmetries of a magnetized system. 
A relativistic system in vacuum is completely symmetric and its energy-momentum tensor is expressed through the only tensor at our disposal, which is in this case the metric tensor $\eta^{\mu\nu}$. In the presence of a medium (i.e. at finite temperature and/or density), the Lorentz symmetry is broken and a new vector (the four velocity of the medium $u_\mu$) gives rise to a new structure, which appears with coefficient $a_2$ in (\ref{Tau}). Now, in the presence of a uniform magnetic field, another symmetry is broken: the rotation symmetry SO(2), and this loss of symmetry has to be reflected in the system's energy-momentum tensor and consequently in the EOS. In this case, there is an extra tensor to form a new structure in $\tau^{\mu\nu}$. It is the electromagnetic strength tensor $F^{\mu\nu}$. Hence, the energy-momentum tensor with the reduced rotational symmetry can be written as a linear combination of three terms. Another way to understand how the magnetic field reduces the system symmetry is by noticing that
because of the breaking of the rotational symmetry the Minkowskian metric is split into two structures one transverse $\eta_{\perp}^{\mu \nu}=\widehat{F}^{\mu \rho}\widehat{F}_{\rho}^\nu$,  and another longitudinal $\eta_{\|}^{\mu \nu}=\eta^{\mu \nu}-\widehat{F}^{\mu \rho}\widehat{F}_{\rho}^\nu$  with respect to the field direction. 

\subsection{Determination of the covariant-structure coefficients}

 To find the $a_i$ coefficients in (\ref{Tau}), a calculation of the quantum-statistical average of the energy-momentum tensor is needed.
\begin{equation} \label{energy-momentum-tensor}
    \big<{\hat{\tau}}^{\mu\nu}\big>=  \frac {  \Tr [ \hat{\tau}^{\mu\nu}e^{ -\beta ( \hat{H}-{\mu}\hat{N}) }] } {\mathcal{Z}}
\end{equation}
where, $\hat{H}$ is the system Hamiltonian operator, $\hat{N}$ is the particle number operator, $\mu$ is the chemical potential and  $\mathcal{Z}$ is the partition function of the grand canonical ensemble, which is given by
\begin{equation} \label{partition function}
    \mathcal{Z}=\Tr[e^{-\beta(\hat{H}-{\mu}\hat{N})}].
\end{equation}

Using functional methods, it was found in \cite{magnetizedfermions} that the coefficients are given by
\begin{equation}\label{Tau2}
\frac{1}{\beta V}\langle\hat{\tau}^{\mu \nu}\rangle=\Omega \eta^{\mu \nu}+(\mu \rho+TS)u^\mu u^\nu +BM \eta_{\perp}^{\mu \nu},
\end{equation}
where $\Omega$ is the system thermodynamic potential, $\rho=-(\partial \Omega / \partial \mu)_{V,T}$ is the average particle-number density, $S=-(\partial \Omega / \partial T)_{V, \mu}$ is the entropy, $M=-(\partial \Omega / \partial B)_{V,T,\mu}$ is the system magnetization, $V$ is the system volume, $B$ is the magnetic field and $\beta=1/T$ is the inverse absolute temperature. 

\subsection{EOS in the presence of a magnetic field}

From (\ref{Tau2}) we can find the system EOS by calculating the different components of $\big<\hat{\tau}^{\mu\nu} \big>$. For a magnetic field directed along the third-spatial direction, the energy density $ \varepsilon$, parallel pressure $P_\parallel$ and  perpendicular  pressure $P_\perp$, can be found from\begin{equation} \label{energy-pressure}
    \varepsilon=\frac{1}{{\beta}V}\big<\hat{\tau}^{00}\big>,\quad P_\parallel=\frac{1}{\beta V}\big<\hat{\tau}^{33}\big>,\quad P_\perp=\frac{1}{{\beta}V}\big<\hat{\tau}^{\perp\perp}\big>
\end{equation}
to be given at $T=0$ by 
\begin{equation} \label{energy-density-1}
    \varepsilon=\Omega+\mu \rho+\frac{B^2}{2},
\end{equation}
\begin{equation} \label{Pressures-EOS-1}
    P_\parallel=-\Omega-\frac{B^2}{2},\quad{P_\perp=-\Omega-MB+\frac{B^2}{2}}.
\end{equation}
In (\ref{energy-density-1})-(\ref{Pressures-EOS-1}) the quadratic terms in $B$ arise from the Maxwell contribution to the energy momentum tensor 
\begin{equation}\label{maxwellset}
	\tau^{\mu\nu}_M=\frac{B^2}{2}\left(\eta^{\mu\nu}_{\parallel}-\eta^{\mu\nu}_{\perp}\right).
\end{equation}

The reason why we are neglecting the temperature effects on the EOS is that for NS in equilibrium $\mu \gg T$, and therefore, the zero-temperature limit is a good approximation. 

It is important to understand the sources of the pressure anisotropy in a magnetized system. First, we have the pure anisotropic effect of the uniform field given by the Maxwell contributions (\ref{maxwellset}), which are proportional to ($\pm B^2$)  and second, the one produced by the system magnetization through the term $BM$. This last term, as was earlier noticed in Ref. \cite{Canuto}, is introduced by the anisotropy of the system energy states. We can model that by considering that the volume occupied by the charged particles in the presence of the magnetic field is given by $V=L(cl_B)^2$, where $L$ is the length along the field direction, $l_B=1/\sqrt{eB}$ is the so called magnetic length and $c$ a numerical coefficient. This expression for the volume takes into account that in the transverse direction the energy states are quantized in Landau orbits with radii proportional to $l_B$. Then, introducing in the partition function the variable changes: $x_3 \rightarrow Lx_3$ and $x_i^\perp \rightarrow c l_B x_i^\perp$  \cite{magnetizedfermions}, we can derive the longitudinal and perpendicular pressures as
\begin{equation} \label{energy-pressure-1}
   P_\parallel=\frac{1}{{\beta}V}\big<\tau^{33}\big>=\frac{1}{{\beta}V}L \frac{d\mathcal{Z}/dL}{\mathcal{Z}} ,                       
\end{equation}

\begin{equation} \label{energy-pressure-2}
  P_\perp=\frac{1}{{\beta}V}\big<\tau^{\perp\perp}\big>=\frac{1}{{\beta}V}l_B \frac{d\mathcal{Z}/dl_B}{\mathcal{Z}}.
\end{equation}

A final essential ingredient to generate the $BM$ term in $P_\perp$ is to take into account that the total derivative in (\ref{energy-pressure-2}) has two terms \cite{magnetizedfermions}

\begin{equation} \label{Tot-Derv}
  \frac{d}{dl_B}=\frac{\partial}{\partial l_B}+ \left (\frac{\partial B}{\partial l_B} \right ) \frac{\partial}{\partial B}=\frac{\partial}{\partial l_B}-2Bl_B^{-1} \frac{\partial}{\partial B}.
\end{equation}
The first term on the right-hand-side of (\ref{Tot-Derv}) measures the effect on the pressure of the variation of the physical volume in the transverse direction, while the second term measures the effect produced by the variation of the effective transverse area occupied by the charged particles under a variation of the magnetic field. From a semiclassical point of view, we can imagine that the field variation produces a redistribution of the Landau orbits, which determine the effective transverse area for the charged particles. If only the first term is considered, then the anisotropy is erased and the system can only be under the Pauli pressure in all directions. Hence, the factor, $BM$, which accounts for the anisotropy in the pressure's transverse direction, manifests when the anisotropy in the energy states is considered. In our approach, it was introduced taking into account the variation of the magnetic length with the magnetic field, while in \cite{Canuto}, it was done by taking the explicit form of the energy states in a second quantized approach.

We should notice that  from the form of the system volume we are considering ($V=L(cl_B)^2$), the corresponding magnetic flux is independent of the magnetic field. This is in agreement with the classical results of London and Onsager \cite{ London} and Saglam and Boyacioglu \cite{M-Flux-Spin} who considered using a non-relativistic semiclassical approach, where the orbit of the charged particle (the electron in their case) is quantized in the presence of the magnetic field, and obtained that the magnetic flux is given as $\Phi=n\Phi_0$,  $n=0,1,2,\cdots$, with $\Phi_0=\frac{2\pi}{e}$ being the so-called quantum fluxoid. Later on in \cite{Magnetic-Flux}, using a quantum relativistic theory, the previous result was validated, showing that the magnetic flux of the charged particles does not depend on the magnetic field. The flux quantization taking place in this system was experimentally confirmed as early as 1961 \cite{Magnetic-Flux-Experiment}.

\subsection{Thermodynamic pressures at $\textbf{B} \neq \textbf{0}$}

To finalize this section, we want to show that the pressures obtained from the quantum-statistical average of the spatial components of the energy-momentum tensor coincide with what is understood as the thermodynamic pressures, which are obtained as the variation of the system's grand canonical potential with respect to the volume.

Let's consider a magnetized system whose grand canonical potential, $\Phi(T, \mu, B, V)$, is a function of temperature, chemical potential, magnetic field and volume. Now we assume that
in the presence of a uniform magnetic field the pressure along the magnetic field direction, $p_\|$, and transverse to that direction, $p_\bot$, are different. Hence, it is natural to expect that the works done by those pressures through their corresponding displacements are in principle not equal. Thus, we differentiate between the variation of the volume along the field direction,  $dV_\|=(L_\bot)^2dL_\|$, to that in the transverse direction, $dV_\bot=2(L_\|L_\bot)dL_\bot$ (Here for the sake of simplicity we change the notation to $L_\bot=cl_B$).

Thus, the grand canonical potential variation is given in this case as
\begin{equation} \label{GCP}
d \Phi=\left (\frac{\partial \Phi}{\partial T}\right )dT+\left (\frac{\partial \Phi}{\partial \mu} \right )d \mu+ \left (\frac{\partial \Phi}{\partial B} \right )dB+ \left (\frac{\partial \Phi}{\partial L_\bot}\right )dL_\bot+ \left (\frac{\partial \Phi}{\partial L_\|}\right )dL_\|.
\end{equation}

As is known, the thermodynamic pressure can be found as the reaction to a virtual change of the effective volume occupied by the particles, and it
 is related to the work done by the system under such a virtual volume deformation. This work depends in general on the interaction affecting the particles in the system. In the case under analysis, we can consider that for the longitudinal deformation of the volume, $dV_\|$, the particle displacements are along the field direction where no magnetic force is acting. This, together with the fact that in our formalism the inter-particle interaction is neglected, implies that ${p}_\|=-(\partial \Phi/ \partial V_\|)$ only depends on the Pauli pressure. For the transverse deformation, $dV_\bot$, the situation is different. Now, the particles' displacements in the transverse direction feel the magnetic interaction. A way to extract the corresponding contribution to the pressure of this magnetic interaction was given in \cite{magnetizedfermions} and discussed above. The guiding idea for this matter is to consider that under a magnetic field the charged particle transverse motion is quantized in Landau orbits  with radii given in units of the magnetic length  $l_B=1/\sqrt{eB}$. 
Thus, to completely determine the work related to the particle displacement in the transverse direction we have to consider the work done by the variation of the effective area available to the charged particles when it is changed by varying the number density of Landau orbits. The density is changed in this case through the variation of the orbits' radii (i.e. by changing $l_B$) while keeping the physical area constant. This contribution is included in the third term of the rhs of (\ref{GCP}) when taking 
\begin{equation} \label{GCP-2}
\left (\frac{\partial \Phi}{\partial B} \right )dB=\left (\frac{\partial \Phi}{\partial B} \right )\left(\frac{\partial l_B}{\partial B} \right )^{-1} dl_B\equiv \left (\frac{\partial \Phi}{\partial B} \right )\left(\frac{\partial L_\perp}{\partial B} \right )^{-1} dL_\perp
\end{equation}

Hence, the expression (\ref{GCP}) can be rewritten as
 \begin{eqnarray} \label{GCP-2}
d \Phi&=&\left (\frac{\partial \Phi}{\partial T}\right )dT+\left (\frac{\partial \Phi}{\partial \mu} \right )d \mu+\left (\frac{\partial \Phi}{L_\perp^2\partial L_\|}\right )dV_\|\nonumber
 \\
 &+&\left [  \left (\frac{\partial \Phi}{2L_\| L_\perp\partial L_\perp}\right )dV_\perp+\left (\frac{\partial \Phi}{\partial B} \right )\left(\frac{\partial L_\perp}{\partial B} \right )^{-1} dL_\perp \right ].
\end{eqnarray}

From where we obtain
\begin{eqnarray} \label{Therm-Pressures-2}
p_\|=-\frac{1}{L_\perp^2}\left (\frac{\partial \Phi}{\partial L_\|}\right )=-\left (\frac{\partial \Phi}{\partial V_\|}\right )=-\Omega,  
\end{eqnarray}

\begin{eqnarray} \label{Therm-Pressures-3}
 p_\bot &=& -\frac{1}{2L_\| L_\perp} \left [  \left (\frac{\partial \Phi}{\partial L_\perp}\right )+\left (\frac{\partial \Phi}{\partial B} \right )\left(\frac{\partial L_\perp}{\partial B} \right )^{-1}  \right ]\nonumber
 \\
&=&-\left (\frac{\partial \Phi}{\partial V_\perp}\right ) +\frac{B}{L_\|(L_\perp)^2}\frac{\partial \Phi}{\partial B}=-\Omega-BM
\end{eqnarray}
Where we took into account that $\Phi=V\Omega$.

Expressions (\ref{Therm-Pressures-2}) and (\ref{Therm-Pressures-3}) coincide with the many-particle contribution of (\ref{Pressures-EOS-1}) (i.e. we obtain (\ref{Pressures-EOS-1}) without the pure Maxwell contribution). Thus, the thermodynamic pressures coincide with those obtained from $\langle{\hat{\tau}}^{\mu \nu}\rangle$, but once again, the contribution of the magnetic field to the variation of the effective volume due to the redistribution of Landau levels has to be taken into account to produce the pressure anisotropy.

\section{First-order phase transition at B $\neq$ 0}

As has been indicated by different calculations \cite{First-order, Greiner} the deconfinement phase transition at high densities is a first-order phase transition. In this section we will analyze how this first-order phase transition is modified in the presence of a magnetic field  the anisotropic EOS when is taken into account.

\subsection{Anisotropic equilibrium conditions at $\textbf{B} \neq \textbf{0}$}

To derive the equilibrium conditions governing the first-order phase transition in this magnetized system, we consider two subsystems $A^{(1)}$ and $A^{(2)}$, representing two phases in contact, such that the total system
\begin{equation}
   A=A^{(1)}+A^{(2)}
  \end{equation}
is isolated. Any process inside $A$ should satisfy the following conditions:

(1) The total internal energy $U^{(A)}=U^{(1)}+U^{(2)}$ remains fixed (i.e. $dU^{(A)}=0$),

(2) The total volume $V^{(A)}=V^{(1)}+V^{(2)}$ remains fixed, so that no work is exchanged with the environment. 

(3) The number of particles of each specie $j$ (i.e. baryons, leptons, etc.) remains fixed inside $A$. Thus $dN^{(A)}_j=dN_j^{(1)}+dN_j^{(2)}=0$.

Now, taking into account the results of the previous section, the differential entropy of system $A$ can be given as the sum of those of the two subsystems
\begin{eqnarray} \label{S-1}
d S_A&=&d S^{(1)}+d S^{(2)} \nonumber
 \\
&=&\frac{dU^{(1)}+ {p}^{(1)}_\| dV_\|^{(1)}+{p}^{(1)}_\bot dV_\bot^{(1)}-\sum_j\mu_j^{(1)} dN_j^{(1)}}{T^{(1)}}\nonumber
 \\
&+&\frac{dU^{(2)}+ {p}_\|^{(2)} dV_\|^{(2)}+{p}_\bot^{(2)} dV_\bot^{(2)}-\sum_j\mu_j^{(2)} dN_j^{(2)}}{T^{(2)}}
\end{eqnarray} 
where $\mu^{(1,2)}$ denotes the chemical potentials of subsystems $A^{(1)}$ and $A^{(2)}$ respectively, $N_j^{(i)}$ the corresponding particle numbers, ${p}_\|^{(i)}$ the thermodynamic parallel pressures and ${p}_\bot^{(i)}$ the transverse pressures, for subsystems $i=1,2$ respectively. Notice that we are separating the longitudinal and transverse volume deformations in agreement with previous analysis.

Eq. (\ref{S-1}) can be rewritten as
\begin{eqnarray} \label{S-2}
d S_A&=&\left ( \frac{dU^{(1)}}{T^{(1)}}+\frac{dU^{(2)}}{T^{(2)}} \right) -\sum_j \left ( \frac{\mu_j^{(1)}}{T^{(1)}} dN_j^{(1)}+\frac{\mu_j^{(2)}}{T^{(2)}} dN_j^{(2)} \right ) \nonumber
 \\
&+&\left (\frac{{p}^{(1)}_\| dV_\|^{(1)} }{T^{(1)}}+\frac{ {p}^{(2)}_\| dV_\|^{(2)} }{T^{(2)}}\right ) +\left (\frac{ {p}^{(1)}_\bot dV_\bot^{(1)}}{T^{(1)}}+\frac{{p}^{(2)}_\bot dV_\bot^{(2)}}{T^{(2)}} \right ).
\end{eqnarray} 

From the isolation conditions we have
\begin{equation} \label{Variation-U}
d U_A=d U^{(1)}+d U^{(2)}=0,
\end{equation}

\begin{equation}\label{Variation-L-par}
d V^{(1)}_\|+d V^{(2)}_\|=0 ,
\end{equation}

\begin{equation}\label{Variation-L-per}
d V^{(1)}_\bot+d V^{(2)}_\bot=0 ,
\end{equation}

\begin{equation}\label{Variation-N}
dN_j^{(A)}= d N_j^{(1)} + d N_j^{(2)}=0.
\end{equation}

Substituting with (\ref{Variation-U})-(\ref{Variation-N}) into (\ref{S-2}), we find
\begin{eqnarray} \label{S-3}
d S_A&=&\left ( \frac{1}{T^{(1)}}-\frac{1}{T^{(2)}} \right) dU^{(1)} -\sum_j\left ( \frac{\mu_j^{(1)}}{T^{(1)}} -\frac{\mu_j^{(2)}}{T^{(2)}} \right )dN_j^{(1)} \nonumber
 \\
&+&\left (\frac{{p}^{(1)}_\| }{T^{(1)}}-\frac{{p}^{(2)}_\|  }{T^{(2)}}\right ) dV_\|^{(1)} +\left (\frac{{p}^{(1)}_\bot }{T^{(1)}}-\frac{{p}^{(2)}_\bot }{T^{(2)}} \right ) dV_\bot^{(1)}.
\end{eqnarray} 

As is well known, the equilibrium condition requires that the entropy is at a maximum and as a consequence, any change in the entropy will be second order in its variables. Hence, there is no linear term at the maximum, so $d S_A=0$.

Therefore, taking into account that $d U_1$, $d V^{(1)}_\|$, $d V^{(1)}_\bot$ and $d N_1$ are independent variations, $d S_A=0$ can only hold if the coefficient of each variation in (\ref{S-3}) vanishes identically. Thus, it follows that
\begin{equation}\label{Thermal-Eq}
T^{(1)}=T^{(2)},
\end{equation}

\begin{equation}\label{Chemical-Eq}
\mu_j^{(1)}=\mu_j^{(2)},
\end{equation}

\begin{equation}\label{Mec-Eq-1}
{p}_\|^{(1)}={p}_\|^{(2)},
\end{equation}

\begin{equation}\label{Mec-Eq-2}
{p}_\bot^{(1)}={p}_\bot^{(2)}.
\end{equation}
Where (\ref{Thermal-Eq}) is the thermal equilibrium condition, (\ref{Chemical-Eq}) is the chemical equilibrium condition and (\ref{Variation-L-par})-(\ref{Variation-L-per}) are the mechanical equilibrium conditions for the anisotropic system.

\subsection{The neutrality condition}

As is well known, symmetries associated with global gauge transformations produce conservations of different charges. For NS astrophysics the conservation of the electric charge and in particular its neutrality is of interest since the stellar matter is considered electrically neutral and in beta equilibrium. Hence, when modeling the inner composition of the star, which is formed by two phases in contact through a phase transition boundary there exist two different constructions:

\textbf{\textit{Gibbs Construction}}

In this case \cite{Rabhi, Glendenning, {Gibbs Cons}} a global neutrality condition is imposed, in such a way that the two coexisting phases have opposite electric charges and a unique electric chemical potential, $\mu_e$. This condition can be written by requiring that the average charge in the mixed phase formed close to the border between the two phases is zero
\begin{equation}\label{Neutrality condition-GC}
	F \frac{\partial\Omega_1}{\partial \mu_e}+(1-F)\frac{\partial \Omega_2}{\partial \mu_e}=0
\end{equation}
Here, $\Omega_{1,2}$ correspond to the thermodynamic potentials of phases $1$ and 2 respectively, and $F=V_1/V$ is the volume fraction of phase 1, which takes values in the interval  $ 0 \leqslant F \leqslant 1$. Thus, this mixed phase is a mixture of the two phases with $F=0$ corresponding to pure phase 1 and $F=1$ to pure phase 2. For $F$ values in between, a mixture of phases takes place occupying different spatial volumes. Thus, this mixed phase will be an inhomogeneous state of matter with domains corresponding to the two different phases \cite{Greiner}.  

\textbf{\textit{Maxwell Construction}}

In this construction \cite{Greiner, MC}, two pure phases are considered in direct contact with each other and each of the phases are independently treated as electrically neutral. In this case, to realize the local neutrality conditions we have to introduce  two chemical potentials, $\mu_e^{(1)}$ for phase 1 and $\mu_2^{(2)}$  for phase 2.
The local neutrality conditions are then written as
\begin{equation}\label{Neutrality condition-MC}
	\frac{\partial \Omega_1}{\partial \mu_e^{(1)}}=0, \quad \quad  \frac{\partial \Omega_2}{\partial \mu_e^{(2)}}=0.
\end{equation}

Moreover, while for the electric chemical potential there is a jump from one phase to the other, in the case of having additional chemical potentials associated to non-gauge global symmetries, such as baryonic or leptonic ones, the chemical equilibrium for them will imply the continuity of the corresponding chemical potentials between the two phases as in (\ref{Chemical-Eq}). 

 We should point out that the Gibbs construction was taken to be the most stable phase when considering more than one chemical potential associated with a system of conserved charges (in our case the baryonic and electric charges) \cite{Glendenning}. The alternative Maxwell construction was then considered unstable due to the possibility for particles to fall from higher energy levels corresponding to the higher electric chemical potential of one phase to the lower energy levels of the other phase. Nevertheless, those conclusions were reached without taking into account the contribution of the surface tension effect and Coulomb interaction across the mixed phase. Once the surface tension ${\lambda}$ is introduced after assuming a sharp boundary between the two phases, one or the other construction may be more appropriate depending on the value of ${\lambda}$ \cite{Surface-Tension}. There are even arguments \cite{MC-Arguments} pointing out that the Gibbs mixed phase is energetically too expensive and that it may be expelled from the stellar medium.

The value of the surface tension of the hadron-quark interface is poorly known. At zero magnetic field there exist some estimates in the range ${\lambda} \approx 10 - 100 MeV/fm^2$ \cite{Sigma-Estimates} and in the presence of a magnetic field in the range of $10^{17}-10^{18}$ G and for the characteristic values of baryon densities ($2n_o - 4n_o$) the values of ${\lambda}$ were found to be between $0.2  MeV/fm^2$ and $15 MeV/fm^2$ \cite{Lugones} . In Ref. \cite{Tatsumi} it was found that at low values of ${\lambda}$ the general construction approaches the Gibbs construction, while for ${\lambda} >{\lambda}_{crit}=60 MeV/fm^2$ it approaches the Maxwell construction. Moreover, it was shown that Gibbs constructions including ${\lambda}$ and/or Coulomb interaction do not give rise to significant differences with respect to the Maxwell construction regarding the bulk properties of compact stars, such as the mass-radius relationship. Nevertheless, due to significant uncertainties in the model parameters, as for instance, the interface energy, the results are yet inconclusive \cite{Greiner}.

Another open question related to finite-size effects that merits investigation is the following. In Ref. \cite{Fukushima}, it was found that the Landau quantization is modified for a magnetized system with cylindrical boundary conditions. This new quantization could in principle alter the thermodynamic potential and consequently the derived quantities, such as magnetization, etc. 

\subsection{Magnetic boundary conditions and magnetic monopoles}

 From the expressions in (\ref{Pressures-EOS-1}), we can rewrite the transverse pressure as
 \begin{equation}\label{B-Jump-4}
p_\perp=p_\|-MB+B^2.
\end{equation}

Then, using (\ref{B-Jump-4}) in  the mechanical equilibrium equations (\ref{Mec-Eq-1}) and (\ref{Mec-Eq-2}) we obtain
\begin{equation}\label{B-Jump-3}
(B_1)^2-B_1M_1=(B_2)^2-B_2M_2.
\end{equation}

From (\ref{B-Jump-3}), a jump in the magnetic field from one phase to the other is implied by the fact that the magnetization, being the first derivative of the thermodynamic potential with respect to the magnetic field,  should be discontinuous along the first-order phase transition boundary. Thus,  if we were to consider that the magnetic field is the same in the two phases, then, from Eq. (\ref{B-Jump-3}), we would have that $M_1=M_2$, which is contradictory. Hence, we should expect that $B_1\neq B_2$.

The jump in the magnetic field around the border of the two phases implies that the divergence of the magnetic field will be  nonzero in that region. Hence, the existence of magnetic monopoles on the phase boundary should be expcted with a density $\rho_M$ given by
 
\begin{equation}\label{Magnetic-Monopole-1}
\bigtriangledown \cdot B=4\pi\rho_M.
\end{equation}

The jump in the magnetic field will be then given by
\begin{equation}\label{Magnetic-Monopole-2}
B_2-B_1=4\pi\sigma_M,
\end{equation}
where if assuming that the stellar magnetic field is in the outward direction,  $B_2$ is the field component perpendicular to the surface separating the two phases immediately above, and $B_1$ is just below the surface. In (\ref{Magnetic-Monopole-2}), $\sigma_M$ is the  monopole surface density on the phase boundary.

Here the following comment is in order to explain how the magnetic monopole charge density can be created in the phase-transition boundary. The creation of magnetic monopoles in the presence of a sufficiently high magnetic field has been already conceived \cite{Manton}. The proposed monopole pair production mechanism proceeds via the magnetic dual of Schwinger pair production \cite{Schwinger}, where as is known, in the presence of an electric field, empty space is unstable to the production of electron-positron pairs. Magnetic monopoles created by stellar magnetic fields, specifically for those of magnetars, have been considered in Ref. \cite{Gould}. Even in Ref \cite{MP-NS}, a method to detect the created magnetic monopoles through the observation of gravitational waves has been proposed. 

Now, to explain the accumulation of magnetic monopoles in the phase-transition boundary, let us assume that the star inner magnetic field on a small patch on the phase boundary is pointing from the star center to its surface and take into account that if a pair of magnetic monopoles are produced in the hadronic phase near the phase-transition boundary, the magnetic field of this phase would pull the magnetic monopole with a negative magnetic charge to the boundary and expel the one with a positive charge toward the surface, while the magnetic field in the quark phase will push the magnetic monopoles with positive magnetic charge to the phase-transition boundary and expel to the star center the monopoles with negative magnetic charge. Since the production rate of monopoles depends on the magnetic-field strength \cite{Manton, MM-Accumulation} and because in our approach the magnitude of the magnetic field performs a jump around the phase transition region, it follows that close to the phase-transition boundary a net accumulation of magnetic charge is created .


\section{Anisotropic Hadron-Quark Phase Transition}

To investigate the hadron-quark  phase transition we consider, for example, that the quark phase occupies the subsystem $A^{(1)}$, while the hadronic phase is in the subsystem $A^{(2)}$, with the quark phase occupying the region closer to the star center where there is a larger density. To study the first-order phase transition we will consider the Maxwell construction, where the set of equations to be solved are:

The mechanical equilibrium conditions (\ref{Mec-Eq-1})-(\ref{Mec-Eq-2}) for the anisotropic system under consideration
\begin{equation}\label{Par-Pres}
	p_{\parallel}^{HP}=p_{\parallel}^{QP},
\end{equation}
\begin{equation}\label{Per-Pres}
	p_{\perp}^{HP}=p_{\perp}^{QP},
\end{equation}
where we are introducing the notation $HP$ and $QP$ to denote variables in the hadronic and quark phases respectively. They are replacing respectively the subindices 2 and 1 used previously. 

The neutrality conditions
\begin{equation}\label{Neutrality-Cond}
\frac{\partial\Omega^{HP}}{\partial \mu^{HP}_e}=0, \quad \quad \frac{\partial\Omega^{QP}}{\partial \mu^{QP}_e}=0,
\end{equation}
which in terms of the particle number densities can be written as
\begin{equation}\label{Neutrality}
\rho_p-\rho_e^{HP}=0, \quad \quad \frac{2}{3}\rho_u-\frac{1}{3}\rho_d-\rho^{QP}_e=0,
\end{equation}
where the subindex $p$ is for proton, $e$ for electron, $u$ and $d$ for the corresponding quark flavors.

On the other hand, the chemical equilibrium condition for the baryonic chemical potential reads 
\begin{equation}\label{ChemPotentials-2}
	\mu^{HP}=\mu^{QP}=\mu,
\end{equation}
which means that we have the same chemical potentials $\mu$ in both phases.

Moreover, we consider that the system is $\beta$-equilibrated, which imposes the following constraints on the chemical potentials of the participating particles
\begin{equation}\label{ChemPotentials}
	\mu_n=\mu,\:\:\mu_p=\mu-\mu^{HP}_e,\:\:\mu_u=\frac{1}{3}\mu-\frac{2}{3}\mu^{QP}_e,\:\:\mu_d=\frac{1}{3}\mu+\frac{1}{3}\mu^{QP}_e.
\end{equation}

Finally, we should impose the minimum equations for the expectation values of the meson fields $\sigma$, $\omega_0$ and $\rho_0$ that participate in the hadronic phase,

\begin{equation}\label{Min-Eq}
\frac{\partial\Omega^{HP}}{\partial\sigma}=\frac{\partial\Omega^{HP}}{\partial\omega_0}=\frac{\partial\Omega^{HP}}{\partial\rho_0}=0.
\end{equation}
These equations give nontrivial expectation values  $\bar{\omega}_0=\left<\omega_0\right>$, $\bar{\rho}_0=\left<\rho^0\right>$ thanks to the nonzero particle number densities. The same is not true for the spatial expectation values of the meson fields, since there is only trivial values for the current densities. 

Therefore, there are eight unknown parameters in the system:  $B^{QP},B^{HP},\mu,\mu^{QP}_e,\mu^{HP}_e,\sigma,\omega_0,\rho_0$ and seven independent equations  given in (\ref{Par-Pres}), (\ref{Per-Pres}), (\ref{Neutrality}) and (\ref{Min-Eq}). Then, by fixing $B^{HP}$ we can find the rest of the parameters. 

In the next section we find the EOS for each phase from where we can solve the phase-transition equations (\ref{Par-Pres}), (\ref{Per-Pres}), (\ref{Neutrality}) and (\ref{Min-Eq}).

We should point out, that as we discussed in Sec. III-B, due to significant uncertainties in the model parameters, which of the two constructions (Maxwell or Gibbs) is more favorable in describing the hadron-quark phase transition has not yet been definitively determined. Thus, in this paper, we arbitrarily selected the Maxwell construction and left the analysis of the anisotropic hadron-quark phase transition in the Gibbs construction for a later study.

\subsection{Hadron Sector Equations of State at $B\neq0$}\label{subsection2a}

Here, $\big<{\hat{\tilde{\tau}}}^{\mu\nu}\big>$ is the quantum-statistical average of the energy momentum tensor given by 
\begin{equation} \label{energy-momentum-tensor}
    \big<{\hat{\tilde{\tau}}}^{\mu\nu}\big>=\frac{\Tr[{\hat{\tilde{\tau}}}^{\mu\nu}e^{-\beta(\hat{H}-{\sum_i\mu_i}\hat{N}_i)}]}{\mathcal{Z}}
\end{equation}
where, $\hat{H}$ is the system Hamiltonian and 
\begin{equation} \label{energy-momentum-tensor-2}
    {\hat{\tilde{\tau}}}^{\mu\nu}=\int_{0}^{\beta}d\tau\int{d^3x[\hat{\tau}^{\mu\nu}_M+\hat{\tau}^{\mu\nu}_f]}
\end{equation}
with $\hat{\tau}^{\mu\nu}_M$ and $\hat{\tau}^{\mu\nu}_f$ being the contributions to the canonically quantized energy-momentum tensor arising from the pure magnetic field (Maxwell contribution) and from the many-particle system respectively, and $\mathcal{Z}$ being the partition function of the grand canonical ensemble, which is given by
\begin{equation} \label{partition function}
    \mathcal{Z}=\Tr[e^{-\beta(\hat{H}-{\sum_i\mu_i}\hat{N}_i)}].
\end{equation}
   
 As we already pointed out in the Introduction, for the HP we will consider the NLW model, where several mesons ($\sigma, \omega_\mu$  and $\rho_\mu$) will mediate the interaction between hadrons. For these boson fields we will show in Appendix A, that in the mean-field approximation, where the meson field components with non-vanishing expectation values are $\bar{\sigma}$, $\bar{\omega}_0$ and $\bar{\rho}_0$, the corresponding statistical average of the corresponding pure meson stress-energy tensor is given by
 \begin{equation}\label{Tau-2}
\frac{1}{\beta V}\langle{\hat{\tilde{\tau}}}_{m}^{\mu \nu}\rangle=\Omega_{m} \eta^{\mu \nu}+ (\bar{\omega}_0{\rho}_\omega+\bar{\rho}_0{\rho}_\rho)  u^\mu u^\nu,
\end{equation}
where $\Omega_m$ is the meson contribution to the thermodynamic potential (see Eq. (\ref{Omega-m})),  ${\rho}_\omega=-\partial\Omega_{m}/\partial\bar{\omega}_0$ and ${\rho}_\rho=-\partial\Omega_{m}/\partial\bar{\rho}_0$. We can see from (\ref{Tau-2}) that $\langle{\hat{\tilde{\tau}}}_{m}^{\mu \nu}\rangle$ is an isotropic stress-energy tensor, which does not give rise to any splitting in the pressure, as was the case with the fermions (\ref{Tau2}). Notice that in (\ref{Tau-2}) the field expectation values $\bar{\omega}_0$ and $\bar{\rho}_0$ play the role of chemical potentials. 

From (\ref{Tau-2}), we have that the meson contribution to the EOS is given by
\begin{equation} \label{EOS-Mesons}
   \varepsilon_m=\Omega_m+(\bar{\omega}_0{\rho}_\omega+\bar{\rho}_0{\rho}_\rho), \quad  P_\parallel=P_\perp=-\Omega_m .
\end{equation}

The meson fields also enter the fermion Lagrangian density through Yukawa couplings. We show in Appendix A that (\ref{Tau2}) is modified such that $\mu\to\mu^*$, where $\mu^*$ is the effective chemical potential, which includes the expectation values $\bar{\omega}_0$ and $\bar{\rho}_0$ (see Eq. (\ref{Chem-Pot}) below).

In what follows, we will derive the corresponding EOS for the magnetized HP taking into consideration the pressure anisotropy created by the magnetic field by itself, as well as by its interaction with the charged fermions.

We start by calculating the thermodynamic potential that describes the HP of hybrid stars. With this goal in mind, we consider the NLW model in a uniform background magnetic field $B$ directed along the $z$-axis. 
The Lagrangian density of this system can be expressed as a sum over its  baryon ($b$), lepton ($l$), meson ($m$), and Maxwell ($M$) components
\begin{equation} \label{LagrangianHadrons-1}
	\mathcal{L}_{HP}=\displaystyle\sum_b\mathcal{L}_b+\displaystyle\sum_l\mathcal{L}_l+\mathcal{L}_m+\mathcal{L}_M,
\end{equation}
where 
\begin{equation} \label{LagrangianHadrons-2}
\begin{aligned}
	  &\mathcal{L}_b=\bar{\psi}_b\left(i\gamma_{\mu}\partial^{\mu}-q_b\gamma_{\mu}A^{\mu}-m_b+g_{\sigma{b}}\sigma-g_{\omega{b}}\gamma_{\mu}\omega^{\mu}-g_{\rho{b}}\tau_{3_b}\gamma_{\mu}\rho^{\mu}\right)\psi_b
\\
& \mathcal{L}_l=\bar{\psi}_l\left(i\gamma_{\mu}\partial^{\mu}-q_l\gamma_{\mu}A^{\mu}-m_l\right)\psi_l
\\
 &\mathcal{L}_m=\frac{1}{2}\partial_{\mu}\sigma\partial^{\mu}\sigma-\frac{1}{2}m_{\sigma}^2\sigma^2-U(\sigma)+\frac{1}{2}m_{\omega}^2\omega_{\mu}\omega^{\mu}-\frac{1}{4}\Omega^{\mu\nu}\Omega_{\mu\nu}+\frac{1}{2}m_{\rho}^2\rho_{\mu}\rho^{\mu}-\frac{1}{4}P^{\mu\nu}P_{\mu\nu}
\\
  &\mathcal{L}_M=-\frac{1}{4}F^{\mu\nu}F_{\mu\nu}.
\end{aligned}
\end{equation}

The baryon index $b$ may include a substantial subset of the lighter baryons, however here we only consider a system with baryon content composed of neutrons and protons $b=n$, $p$. Similarly, we take the lepton content to be composed only of electrons, $l=e$. In (\ref{LagrangianHadrons-2}), $\tau_{3_b}$ denotes the baryon isospin projection operator, $g_{\sigma b}$, $g_{\omega b}$ and $g_{\rho b}$ are the baryon-meson couplings (see Table I \cite{Table}), and the mesonic and electromagnetic field tensors are given by: $\Omega_{\mu\nu}=\partial_{\mu}\omega_{\nu}-\partial_{\nu}\omega_{\mu}$, $P_{\mu\nu}=\partial_{\mu}\rho_{\nu}-\partial_{\nu}\rho_{\mu}$, and $F_{\mu\nu}=\partial_{\mu}A_{\nu}-\partial_{\nu}A_{\mu}$. 

The scalar self-interaction potential is (see Table \ref{parameters} for a list of parameter values).
\begin{equation}\label{Potential}
	U(\sigma)=\frac{1}{3}cm_n(g_{\sigma{N}}\sigma)^3+\frac{1}{4}d(g_{\sigma{N}}\sigma)^4.
\end{equation}

 In \cite{insignificance}  it was found that the magnetic field-anomalous magnetic moment (B-AMM) interaction does not significantly affect the EOS of charged fermions and in \cite{neutrons} it was found that the B-AMM interaction is insignificant for neutrons at magnetic field strengths less than $10^{18}$G. We omit the B-AMM interaction terms in (\ref{LagrangianHadrons-2}), however the effects of the neutron B-AMM on the hadron-quark phase transition at magnetic fields of order $10^{18}$ G will be studied later on in the paper. 
\begin{table}[ht]
\caption{Scalar, $g_{\sigma{N}}$, and vector meson-nucleon, $g_{\omega{N}}$, $g_{\rho{N}}$, couplings as well as meson self interaction coefficients, $c$, $d$,  chosen to reproduce the binding energy, baryon density, symmetry energy coefficient and effective mass at nuclear saturation for a compression modulus $K=30$, as reported in Ref. \cite{Rabhi, Table}.}
\centering
  \begin{tabular}{ | m{1cm}| m{1cm} | m{1cm} |m{1.5cm} |m{1.5cm} | } 
    \hline
 $g_{\sigma{N}}$ & $g_{\omega{N}}$  & $g_{\rho{N}}$ & c & d  \\ 
\hline
 8.910 & 10.610 & 8.196 & .002947 & -0.001070 \\ 
  &  & & &  \\ 
\hline
  \end{tabular}
  \label{parameters}
\end{table}

In the NLW model, mesons are taken in the mean-field approximation, where only their field expectation values $\bar{\sigma}=\left<\sigma\right>$, $\bar{\omega}_0=\left<\omega_0\right>$, $\bar{\rho}_0=\left<\rho^0\right>$ contribute into the one-loop thermodynamic potential. 

The fermion Lagrangian densities (\ref{LagrangianHadrons-2}) admit global symmetries, which have associated conserved number densities  given by
\begin{equation}
	\rho_i=\bar{\psi}_i\gamma^0\psi_i, \quad i=p, n, e
\end{equation}

These conserved number densities enter the partition function (\ref{partition function}) through the relation $\sum_i\mu_i\rho_i$, where the chemical potentials $\mu_i$ play the role of Lagrange multiplayers.

The fermion contribution to the thermodynamic potential in the HP is given by 
\begin{equation}
	\Omega_f^{HP}=\Omega_n+\Omega_p+\Omega_e,
\end{equation}
where in the one-loop approximation we have

\begin{equation} \label{ThermoPotential-1}
\begin{aligned}
	  &\Omega_n=-\frac{1}{\beta}\displaystyle\sum_{p_4}\int_{-\infty}^{\infty}\frac{d^3p}{(2\pi)^3}\ln\det\left(\slashed{p}^*_n+m_n^*\right)
\\
& \Omega_p=-\frac{eB}{\beta}\displaystyle\sum_{p_4}\int_{-\infty}^{\infty}\frac{dp_3}{(2\pi)^2}\left\{\frac{1}{2}\ln\det\left[\bar{\slashed{p}}^*_{p}(l=0)-m_p^*\right]+\displaystyle\sum_{l=1}^{\infty}\ln\det\left[\bar{\slashed{p}}^*_{p}(l)-m_p^*\right]\right\}
\\
 &\Omega_e=-\frac{eB}{\beta}\displaystyle\sum_{p_4}\int_{-\infty}^{\infty}\frac{dp_3}{(2\pi)^2}\left\{\frac{1}{2}\ln\det\left[\bar{\slashed{p}}^*_{e}(l=0)-m_e\right]+\displaystyle\sum_{l=1}^{\infty}\ln\det\left[\bar{\slashed{p}}^*_{e}(l)-m_e\right]\right\}.
\end{aligned}
\end{equation}

Here, $p_4=\frac{\left(2n+1\right)\pi}{\beta}$ for $n=\left(n=0,\pm1,\pm2,...\right)$, are the Matsubara frequencies for fermions. The four momenta and effective masses for the different particles are

\begin{equation} \label{Momenta-Masses}
\begin{aligned}
	  &p^*_n=\left(p^1, p^2, p^3, {p^4}+i\mu_n-ig_{\omega{N}}\bar{\omega}^0-ig_{\rho{N}}\tau_{3_n}\bar{\rho}^0\right),
\\
&\bar{p}^*_p=(0, -\sqrt{2eBl}, p^3, {p^4}+i\mu_p-ig_{\omega{N}}\bar{\omega}^0-ig_{\rho{N}}\tau_{3_p}\bar{\rho}^0),
 \\
 &\bar{p}^*_e=(0, -\sqrt{2eBl}, p^3, {p^4}+i\mu_e^{HP}), 
  \\
  &m_n^*=m_n-g_{\sigma{N}}\bar{\sigma},
   \\
 &m_p^*=m_p-g_{\sigma{N}}\bar{\sigma}.
\end{aligned}
\end{equation}

Note that a factor of $1/2$ appears in front of the lowest Landau level (LLL) contributions to the charged fermion thermodynamic potentials in (\ref{ThermoPotential-1}) since in the LLL only one spin projection contributes (i.e. there is no spin degeneracy.)

Performing the sums in Matsubara frequencies and calculating the determinants in (\ref{ThermoPotential-1}), we obtain the one-loop fermion thermodynamic potential as a sum of its vacuum (QFT) and finite temperature-statistical (S) components.
\begin{equation}
	\Omega_f^{HP}=\Omega_n^{QFT}+\Omega_n^{S}+\Omega_p^{QFT}+\Omega_p^{S}+\Omega_e^{QFT}+\Omega_e^{S},
\end{equation}
where
\begin{equation}\label{ThermoPotentialVac}
\begin{aligned}
	&\Omega_n^{QFT}=-2\int_{-\infty}^{\infty}\frac{d^3p}{(2\pi)^3}E_n,
\\
&\Omega_p^{QFT}=-eB\int_{-\infty}^{\infty}\frac{dp_3}{(2\pi)^2}\left\{E_p\left(l=0\right)+2\displaystyle\sum_{l=1}^{\infty}E_p\right\},
\\
&\Omega_e^{QFT}=-eB\int_{-\infty}^{\infty}\frac{dp_3}{(2\pi)^2}\left\{E_e\left(l=0\right)+2\displaystyle\sum_{l=1}^{\infty}E_e\right\},
\end{aligned}
\end{equation}

and

\begin{equation} \label{ThermoPotentialStat}
\begin{aligned}
	  &\Omega_n^S=-2\int_{-\infty}^{\infty}\frac{d^3p}{(2\pi)^3}\left\{\frac{1}{\beta}\ln\left[1+e^{-\beta\left(E_n+\mu_n^*\right)}\right]+\frac{1}{\beta}\ln\left[1+e^{-\beta\left(E_n-\mu_n^*\right)}\right]\right\},
\\
&\Omega_p^S =-eB\int_{-\infty}^{\infty}\frac{dp_3}{(2\pi)^2}\left\{\frac{1}{\beta}\ln\left[1+e^{-\beta\left(E_p\left(l=0\right)+\mu_p^*\right)}\right]+\frac{1}{\beta}\ln\left[1+e^{-\beta\left(E_p\left(l=0\right)-\mu_p^*\right)}\right]\right\}
\\
& -2eB\displaystyle\sum_{l=1}^{\infty}\int_{-\infty}^{\infty}\frac{dp_3}{(2\pi)^2}\left\{\frac{1}{\beta}\ln\left[1+e^{-\beta\left(E_p+\mu_p^*\right)}\right]+\frac{1}{\beta}\ln\left[1+e^{-\beta\left(E_p-\mu_p^*\right)}\right]\right\},
\\
&\Omega_e^S= -eB\int_{-\infty}^{\infty}\frac{dp_3}{(2\pi)^2}\left\{\frac{1}{\beta}\ln\left[1+e^{-\beta\left(E_e\left(l=0\right)+\mu_e^{HP}\right)}\right]+\frac{1}{\beta}\ln\left[1+e^{-\beta\left(E_e\left(l=0\right)-\mu^{HP}_e\right)}\right]\right\}
\\
& -2eB\displaystyle\sum_{l=1}^{\infty}\int_{-\infty}^{\infty}\frac{dp_3}{(2\pi)^2}\left\{\frac{1}{\beta}\ln\left[1+e^{-\beta\left(E_e+\mu^{HP}_e\right)}\right]+\frac{1}{\beta}\ln\left[1+e^{-\beta\left(E_e-\mu^{HP}_e\right)}\right]\right\}.
\end{aligned}
\end{equation}

Here 
\begin{equation} \label{Chem-Pot}
	\mu_n^*=\mu_n-g_{\omega{N}}{\bar{\omega}^0}-g_{\rho{N}}\tau_{3_n}\bar{\rho}^0,  \quad \mu_p^*=\mu_p-g_{\omega{N}}\bar{\omega}^0-g_{\rho{N}}\tau_{3_p}\bar{\rho}^0,
\end{equation}
and $E_n$, $E_p$, and $E_e$ are given by 
\begin{equation} \label{Disp-Rel}
	E_n=\sqrt{p_1^2+p_2^2+p_3^2+{m_n^*}^2},\:\:E_p=\sqrt{p_3^2+2eBl+{m_p^*}^2},\:\:E_e=\sqrt{p_3^2+2eBl+{m_e}^2}.
\end{equation}

Since for NS the high particle density makes the chemical potentials the leading parameters, it is natural to expect the many-particle contribution to provide the largest contribution to the thermodynamic potential. Dropping the vacuum terms and taking the zero-temperature limit we arrive at the many-particle-fermion thermodynamic potential 
\begin{equation}\label{Omega-f}
	{\Omega_{f}^{\mu}}^{HP}=\Omega_n^{\mu}+\Omega_p^{\mu}+\Omega_e^{\mu},
\end{equation}
where
\begin{equation} \label{ThermoPotentialFiniteDensity}
\begin{aligned}
	  &\Omega_n^{\mu}=-2\int_{-\infty}^{\infty}\frac{d^3p}{(2\pi)^3}\left(\mu_n^*-E_n\right)\Theta\left(\mu_n^*-E_n\right),
\\
&\Omega_p^{\mu} =-eB\int_{-\infty}^{\infty}\frac{dp_3}{(2\pi)^2}\left\{\left(\mu_p^*-E_p(l=0)\right)\Theta\left(\mu_p^*-E_p(l=0)\right)+2\displaystyle\sum_{l=1}^{\infty}\left(\mu_p^*-E_p\right)\Theta\left(\mu_p^*-E_p\right)\right\},
\\
&\Omega_e^{\mu} =-eB\int_{-\infty}^{\infty}\frac{dp_3}{(2\pi)^2}\left\{\left(\mu_e^{HP}-E_e(l=0)\right)\Theta\left(\mu_e^{HP}-E_e(l=0)\right)+2\displaystyle\sum_{l=1}^{\infty}\left(\mu_e^{HP}-E_e\right)\Theta\left(\mu_e^{HP}-E_e\right)\right\}.
\end{aligned}
\end{equation}

It is also expected that $eB< \mu_n^2$; hence, we can limit our calculation to the weak-field approximation (WFA). Carrying out the integration over momentum and using the WFA to approximate the sum over Landau levels (see Appendix \ref{Appendix-B} for details on carrying out the WFA), we arrive at 
\begin{equation} \label{ThermoPotentialFiniteDensity-WFA}
\begin{aligned}
	  &\Omega_n^{\mu}\approx\frac{-1}{24\pi^2}\left\{\left(2{\mu^*_n}^4-5{m_n^*}^2{\mu_n^*}^2\right)\sqrt{1-\left(\frac{m_n^*}{\mu_n^*}\right)^2}+3{m_n^*}^4\ln\left[\frac{{\mu_n^*}+\sqrt{{\mu_n^*}^2-{m_n^*}^2}}{{m_n^*}}\right]\right\},
\\
&\Omega_p^{\mu} \approx\frac{-1}{24\pi^2}\left\{\left(2{\mu^*_p}^4-5{m_p^*}^2{\mu_p^*}^2\right)\sqrt{1-\left(\frac{m_p^*}{\mu_p^*}\right)^2}+\left(3{m_p^*}^4+2\left(eB\right)^2\right)\ln\left[\frac{{\mu_p^*}+\sqrt{{\mu_p^*}^2-{m_p^*}^2}}{{m_p^*}}\right]\right\},
\\
&\Omega_e^{\mu} \approx\frac{-1}{24\pi^2}\left\{\left(2\left({\mu_e^{HP}}\right)^4-5{m_e}^2\left({\mu_e^{HP}}\right)^2\right)\sqrt{1-\left(\frac{m_e}{\mu_e^{HP}}\right)^2}+\left(3{m_e}^4+2\left(eB\right)^2\right)\ln\left[\frac{{\mu_e^{HP}}+\sqrt{\left({\mu_e^{HP}}\right)^2-{m_e}^2}}{{m_e}}\right]\right\}.
\end{aligned}
\end{equation}

Plugging (\ref{Omega-f}) together with (\ref{ThermoPotentialFiniteDensity-WFA}) into (\ref{energy-density-1})-(\ref{Pressures-EOS-1}), taking into account that $\mu_b\to\mu_b^*$ in the energy density expression according to $(\ref{Chem-Pot})$, and adding the pure meson contribution (\ref{EOS-Mesons}), we obtain for the EOS of the hadronic phase
\begin{equation}\label{PressuresHS}
	\begin{aligned}
	&\varepsilon={\Omega_f^{\mu}}^{HP}+\displaystyle\sum_{i}\mu_i^*{\rho}_i^{{\mu}{HP}}+\frac{B^2}{2}+\left\{\frac{1}{2}m_{\sigma}^2\bar{\sigma}^2+U(\bar{\sigma})+\frac{1}{2}m_{\omega}^2{\bar{\omega}_0}^2+\frac{1}{2}m_{\rho}^2{\bar{\rho}_0}^2\right\},
\\
&P_{\perp}=-{\Omega_f^{\mu}}^{HP}-BM_f^{{\mu}HP}+\frac{B^2}{2}+\left\{-\frac{1}{2}m_{\sigma}^2\bar{\sigma}^2-U(\bar{\sigma})+\frac{1}{2}m_{\omega}^2{\bar{\omega}_0}^2+\frac{1}{2}m_{\rho}^2{\bar{\rho}_0}^2\right\},
\\
&P_{\parallel}=-{\Omega_f^{\mu}}^{HP}-\frac{B^2}{2}+\left\{-\frac{1}{2}m_{\sigma}^2\bar{\sigma}^2-U(\bar{\sigma})+\frac{1}{2}m_{\omega}^2{\bar{\omega}_0}^2+\frac{1}{2}m_{\rho}^2{\bar{\rho}_0}^2\right\},
	\end{aligned}
\end{equation}
where the index $i$ runs over fermion species, $M^{\mu HP}_f=-\partial \Omega_f^{\mu HP}/\partial{B}$ is the zero temperature-finite density fermion magnetization,  and $\rho_i^{\mu HP}=-\partial{\Omega}^{\mu HP}/\partial{\mu_i}$ are the $i^{th}$ particle species number densities at zero temperature and finite density, which are given by

\begin{equation}\label{numberdensitiesHS}
\begin{aligned}
	{\rho}^{{\mu}}_n&=\frac{1}{3\pi^2}\left({\mu^*_n}^2-{m^*_n}^2\right)^{\frac{3}{2}},
\\
	{\rho}^{{\mu}}_p&=\frac{1}{3\pi^2}\left[\left({\mu^*_p}^2-{m^*_p}^2\right)^{\frac{3}{2}}+\frac{(eB)^2}{4\sqrt{{\mu^*_p}^2-{m^*_p}^2}}\right],
\\
	{\rho}^{{\mu}}_e&=\frac{1}{3\pi^2}\left[\left(\left({\mu_e^{HP}}\right)^2-m_e^2\right)^{\frac{3}{2}}+\frac{(eB)^2}{4\sqrt{\left({\mu_e^{HP}}\right)^2-m_e^2}}\right],
\end{aligned}
\end{equation}
and the baryonic charge density in the hadron phase is given by
\begin{equation}
	\rho_b^{HP}=\displaystyle\sum_iq^{HP}_i\rho_i^{HP}=\rho^{\mu}_n+\rho^{\mu}_p,
\end{equation}
where $q_i^{HP}$ is the baryonic charge of the i'th particle species in the hadron phase.

\subsection{Quark Sector Equations of State at $B\neq0$}\label{subsection2b}

We consider that the QP is formed by quarks of two flavors and electrons. To describe the quark content of the QP we employ the MIT bag model where the quarks are considered to be free inside of an effective bag, which is realized in the theory by adding (subtracting) a fixed constant to the energy density (pressure), which accounts for the inward pressure needed to confine the quarks into the bag.
The electron content and pure magnetic field contribution are described by the same Lagrangians used to describe the lepton and Maxwell contributions in the hadron sector, keeping in mind that the magnetic field and electric chemical potential may differ in the QP. The QP Lagrangian is given by
\begin{equation}\label{LagrangianQuarks-1}
	\mathcal{L}_{QP}=\displaystyle\sum_q\mathcal{L}_q+\mathcal{L}_e+\mathcal{L}_M,
\end{equation}
where
\begin{equation}\label{LagrangianQuarks-2}
	\mathcal{L}_q=\bar{\psi}_q\left(i\gamma_{\mu}\partial^{\mu}-q_q\gamma_{\mu}A^{\mu}-m_q\right)\psi_q.
\end{equation}

The quark index $q$ runs over up and down flavors, $u,d$ respectively, and is degenerate in the three color charges. We use $m_u=m_d=5.5$MeV. The quark Lagrangian densities also admit global symmetries, which give rise to conserved number densities given by 
\begin{equation}
	{\rho_q}=\bar{\psi}_q\gamma^0\psi_q,
\end{equation}

To each conserved number density a chemical potentials $\mu_q$ can be introduced as a Lagrange multiplayer.

\begin{figure}[t]
\begin{center}
\includegraphics[width=0.5\textwidth]{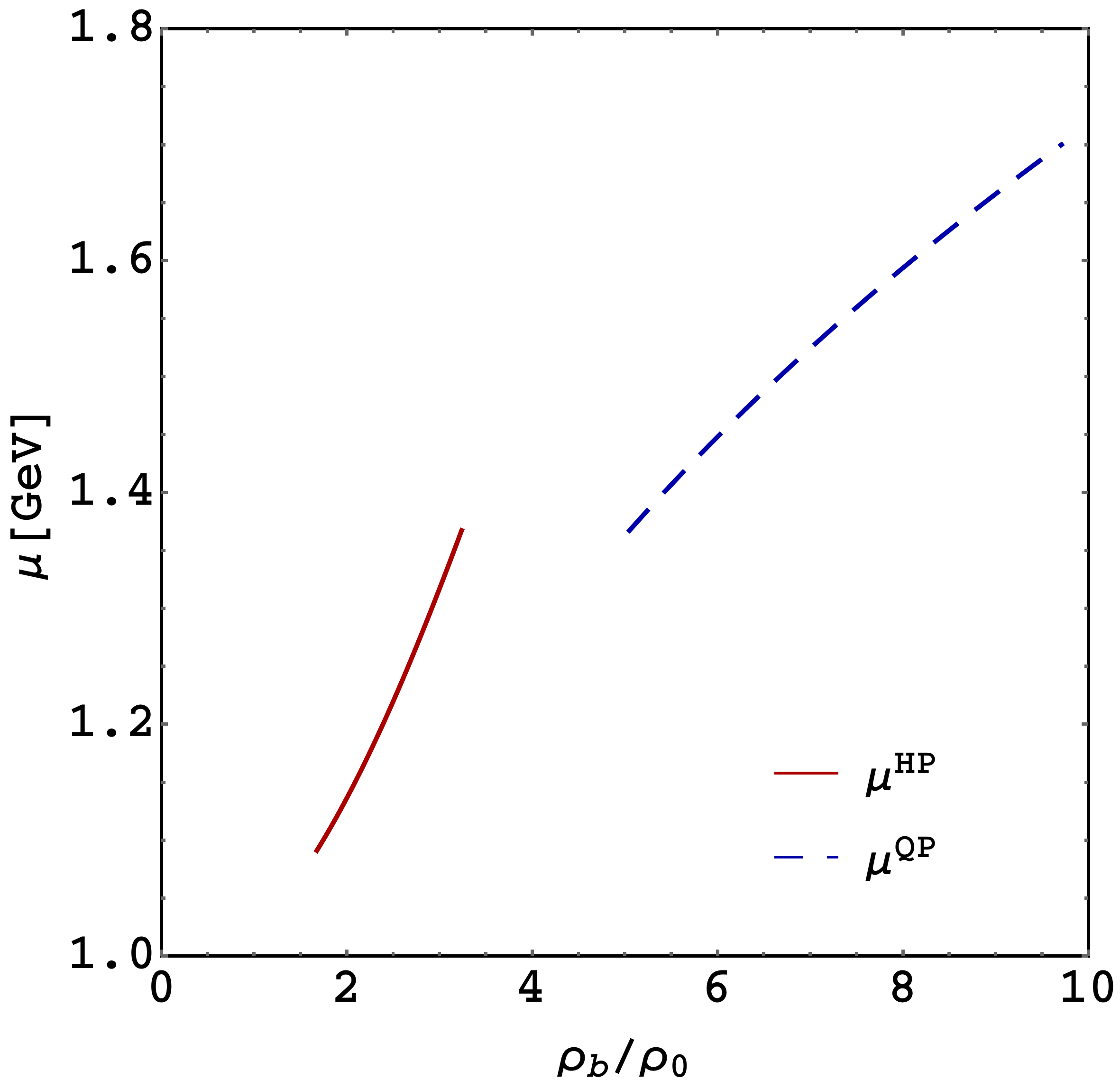}
\caption{(Color online) Baryonic chemical potential, $\mu$ in the HP (red-solid) and the QP (blue-dashed), versus baryonic charge density, $\rho_b$ normalized against the baryonic saturation density $\rho_0=0.153$ fm$^{-3}$, at a magnetic field $B^{HP}=10^{16}$ G. The jump in the baryon number density is signaling the hadron-quark phase transition at $\mu_c=1.367$GeV. }
\label{mubVrho}
\end{center}
\end{figure}

 In the WFA, the finite-density thermodynamic potential for each quark is arrived at by taking $m_e{\to}m_q$, $e\to{q_q}$, $\mu^{HP}_e\to{\mu^{QP}_q}$ in $\Omega_e^{\mu}$, and multiplying by an overall factor of $3$ to account for the color degeneracy. We then have
\begin{equation}\label{ThermoPotentialQuarks-1}
	{\Omega_f^{\mu}}^{QP}=\Omega_u^{\mu}+\Omega_d^{\mu}+\Omega_e^{\mu}
\end{equation}

\begin{equation}\label{ThermoPotentialQuarks-2}
	\begin{aligned}
	  &\Omega_u^{\mu} \approx\frac{-1}{8\pi^2}\left\{\left(2{\mu_u}^4-5{m_u}^2{\mu_u}^2\right)\sqrt{1-\left(\frac{m_u}{\mu_u}\right)^2}+\left(3{m_u}^4+\frac{8}{9}\left(eB\right)^2\right)\ln\left[\frac{{\mu_u}+\sqrt{{\mu_u}^2-{m_u}^2}}{{m_u}}\right]\right\},
\\
&\Omega_d^{\mu} \approx\frac{-1}{8\pi^2}\left\{\left(2{\mu_d}^4-5{m_d}^2{\mu_d}^2\right)\sqrt{1-\left(\frac{m_d}{\mu_d}\right)^2}+\left(3{m_d}^4+\frac{2}{9}\left(eB\right)^2\right)\ln\left[\frac{{\mu_d}+\sqrt{{\mu_d}^2-{m_d}^2}}{{m_d}}\right]\right\},
\\
&\Omega_e^{\mu} \approx\frac{-1}{24\pi^2}\left\{\left(2{(\mu^{QP}_e})^4-5{m_e}^2({\mu^{QP}_e})^2\right)\sqrt{1-\left(\frac{m_e}{\mu^{QP}_e}\right)^2}+\left(3{m_e}^4+2\left(eB\right)^2\right)\ln\left[\frac{{\mu^{QP}_e}+\sqrt{({\mu^{QP}_e})^2-{m_e}^2}}{{m_e}}\right]\right\}.
\end{aligned}
\end{equation}

The EOS for the QP are determined the same way as was done for the HP except here the meson field contribution (\ref{EOS-Mesons}) is removed and the bag constant is added to and subtracted from the energy density and pressures respectively.
\begin{equation}\label{PressuresQS}
	\begin{aligned}
	\varepsilon&={\Omega_f^{\mu}}^{QP}+\displaystyle\sum_{i=u,d,e}\mu_i{\rho}_i^{{{\mu}}QP}+Bag,
\\
P^{QP}_{\perp}&=-{\Omega_f^{\mu}}^{QP}-BM_f^{{{\mu}}QP}+\frac{B^2}{2}-Bag,
\\
P^{QP}_{\parallel}&=-{\Omega_f^{\mu}}^{QP}-\frac{B^2}{2}-Bag.
	\end{aligned}
\end{equation}

The quark number densities are given by
\begin{equation}\label{numberdensitiesQS}
\begin{aligned}
	{\rho}^{{\mu}}_u&=\frac{1}{\pi^2}\left[\left({\mu_u}^2-m_u^2\right)^{\frac{3}{2}}+\frac{(eB)^2}{9\sqrt{{\mu_u}^2-m_u^2}}\right],
\\
	{\rho}^{{\mu}}_d&=\frac{1}{\pi^2}\left[\left({\mu_d}^2-m_d^2\right)^{\frac{3}{2}}+\frac{(eB)^2}{36\sqrt{{\mu_d}^2-m_d^2}}\right].
\end{aligned}
\end{equation}
and the electron number density is similar to that in (\ref{numberdensitiesHS}). The quark phase baryonic charge density is given by 

\begin{equation}
	\rho^{QP}_b=\frac{1}{3}\rho_u+\frac{1}{3}\rho_d.
\end{equation}

In the following, for the numerical calculations we will take $Bag=$(180MeV)$^4$ to be the value of the bag constant.

\subsection{Numerical Solutions for the Anisotropic Quark-Hadron Phase Transition at $\textbf{B} \neq \textbf{0}$ in the Maxwell construction}\label{EQ}\label{section3}

In this section, we draw our attention to the inner region of a NS, which has its hadronic phase closer to its surface. When the radial distance is decreased, it is expected that the matter density increases and eventually, if the density increases enough, a first-order phase transition occurs toward a quark phase. Our goal now is to determine, through Eqs. (\ref{Par-Pres}), (\ref{Per-Pres}), (\ref{Neutrality}) and (\ref{Min-Eq}), the critical values of the parameters that characterize the first-order phase transition in the Maxwell construction.

In Fig. \ref{mubVrho},  by fixing the magnetic-field value in the hadronic phase, $B^{HP}=10^{16}$ G, we determine the relation between the baryonic chemical potential and the baryonic charge density. There, the solid line corresponds to the hadronic phase and the dashed line to the quark phase. By increasing $\mu$ we reach the phase-transition point separating  the hadronic phase from the quark phase at the critical value $\mu_c=1.367$GeV. It can be noticed how the baryonic density jumps at that critical value indicating the first-order nature of the phase transition. 

\begin{figure}
\begin{center}
\begin{tabular}{ccc}
  \includegraphics[width=7cm]{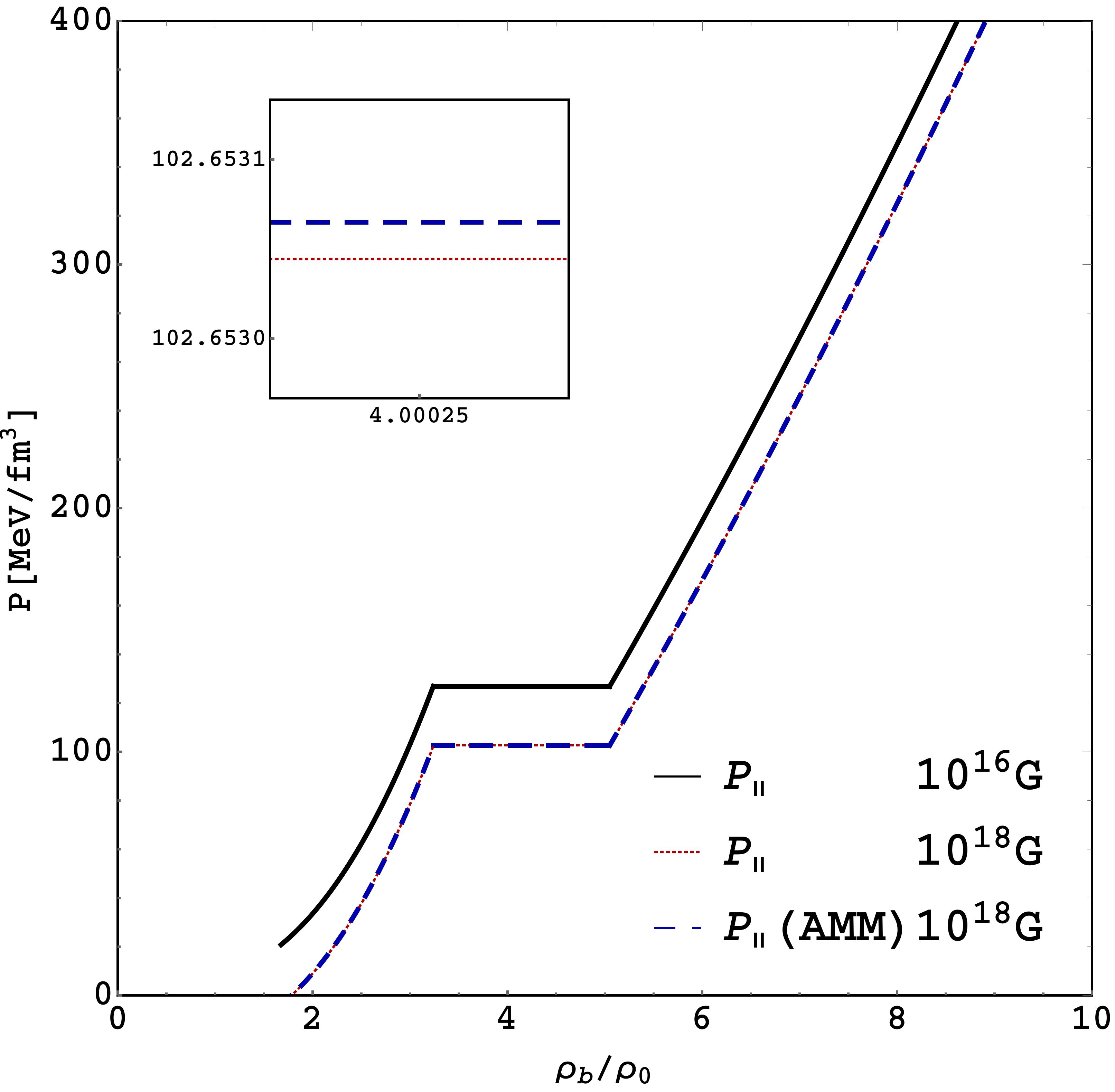} & \includegraphics[width=7cm]{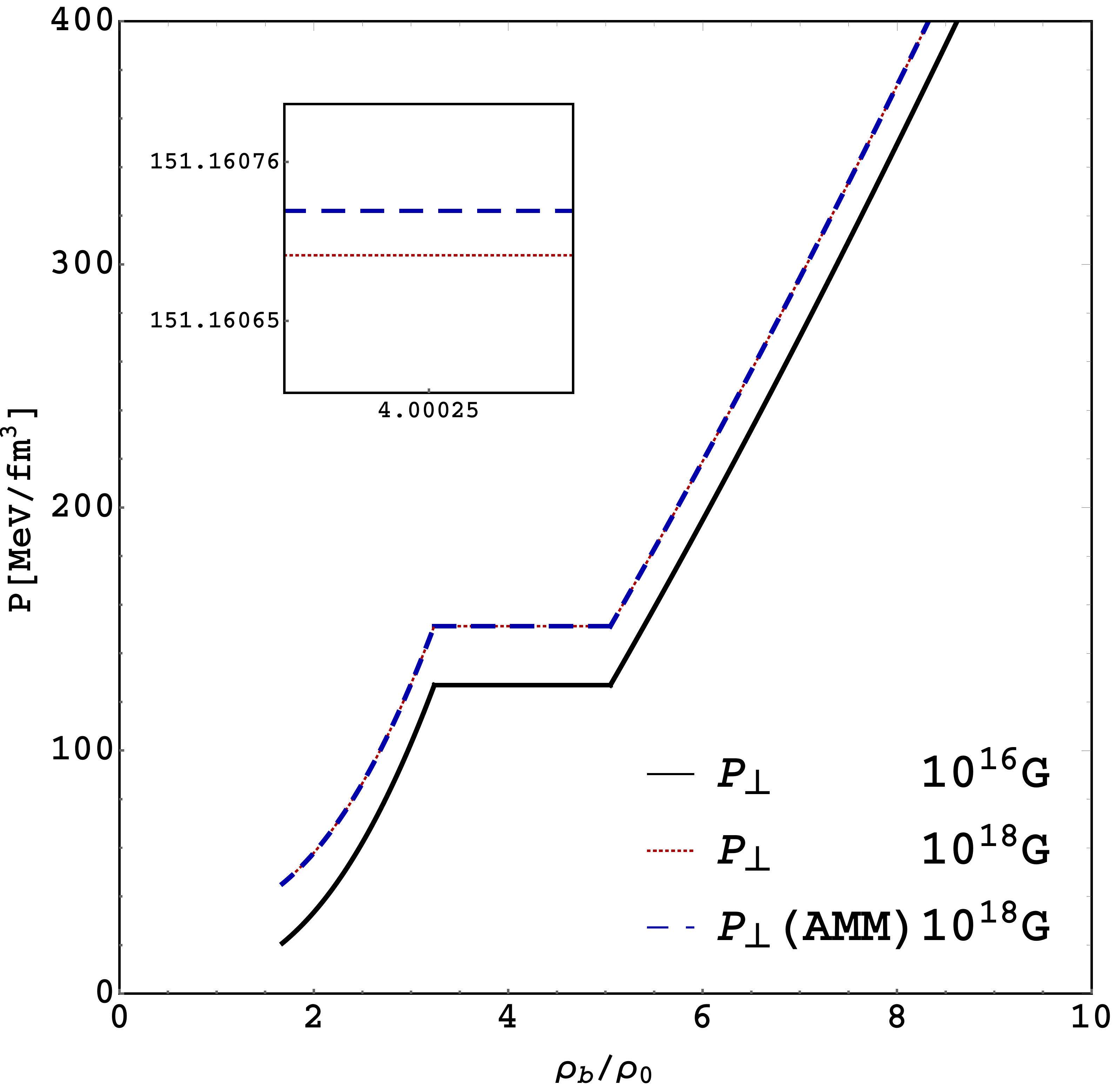}\\
(a) & (b)  \\
  \\
  \end{tabular}
    \end{center}
    \caption{(Color online) Parallel (a) and perpendicular (b) pressures in the Maxwell construction versus baryonic charge density $\rho_b$, normalized against the baryonic saturation density $\rho_0=0.153$ fm$^{-3}$,
    including (blue-dashed) the neutron AMM at $B^{HP}=10^{18}$G and excluding the neutron AMM at $B^{HP}=10^{18}$G (red-dotted) and at $B^{HP}=10^{16}$G (black-solid).}
     \label{Pressures}
\end{figure}

In Fig. \ref{Pressures}, we plot the parallel and perpendicular pressures  without considering  the contribution of the B-AMM interaction of the contributing particles versus baryonic charge density $\rho_b$ normalized against the baryonic saturation density $\rho_0$ for $B^{HP}=10^{16}$ G (solid line) and  for $B^{HP}=10^{18}$ G (dotted line). Then, the pressures are plotted with a dashed line for the case when the B-AMM interaction for neutrons is included at $B^{HP}=10^{18}$ G. There, while the baryon density is discontinuous around the phase-transition point, the pressures are continuous, which is typical for these two magnitudes in a first-order phase transition. 

While in \cite{insignificance} it was shown that the B-AMM contribution of charged particles to the pressures is insignificant up to field of the order of $10^{18}$ G, in \cite{Broderick} it was found that the B-AMM interaction significantly affects the pressure contribution of neutrons at field strengths around $10^{18}$G when the pure Maxwell contribution is set aside. Thus, it remains to determine if the neutron B-AMM interaction is influential on the system's behavior when the pure Maxwell term is also included. In order to determine the effect of the neutron B-AMM interaction on the phase transition in the Maxwell construction the parallel and perpendicular pressures are plotted in Fig. \ref{Pressures} while including and excluding the B-AMM interaction. The neutron component of the free energy with the B-AMM included was taken from \cite{neutrons}. As displayed in Fig. \ref{Pressures}, the B-AMM provides a relatively small influence on the pressures when compared with the effects arising from increasing the magnetic field. 

\begin{figure}
\begin{center}
\begin{tabular}{ccc}
  \includegraphics[width=7cm]{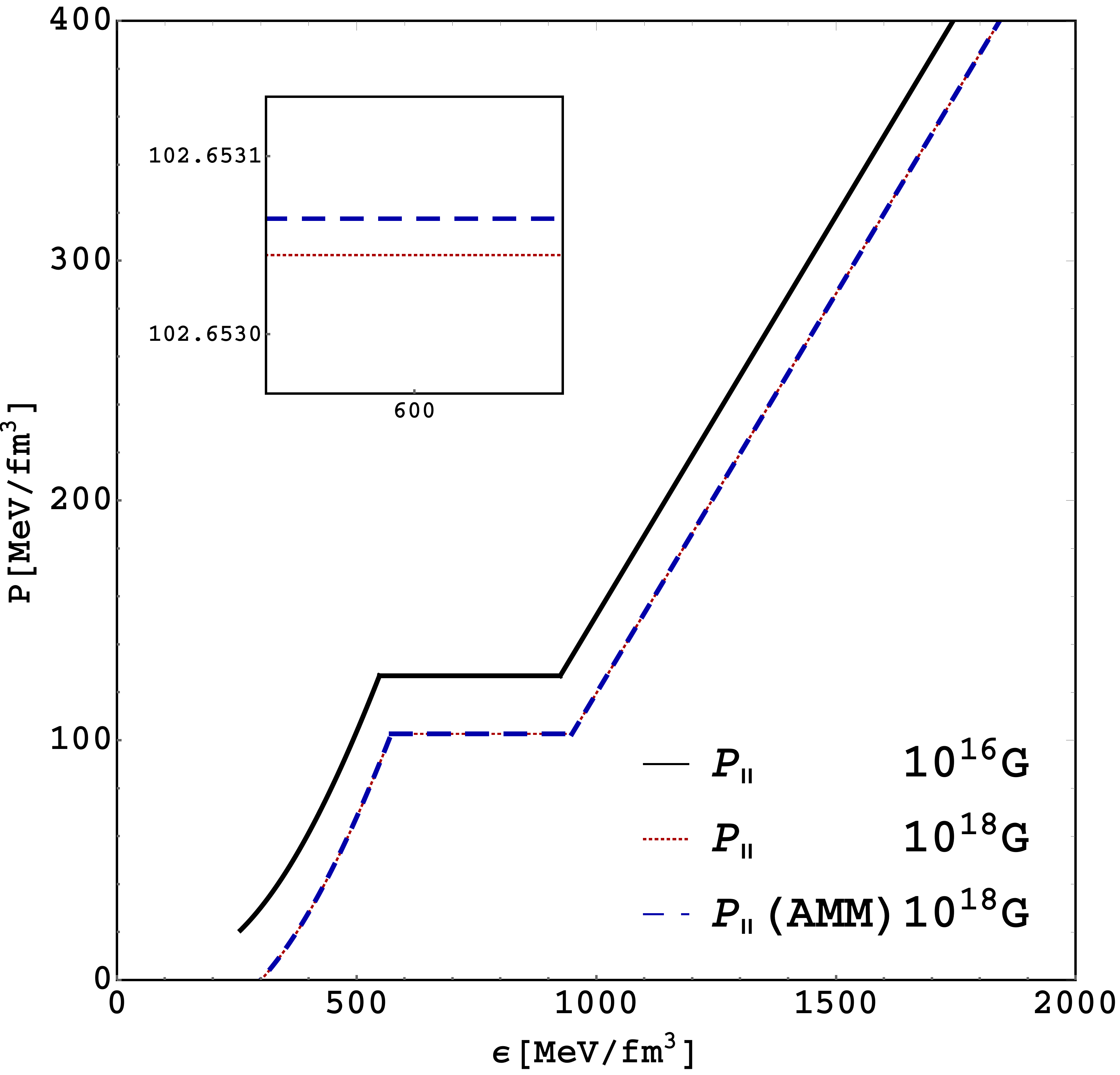} & \includegraphics[width=7cm]{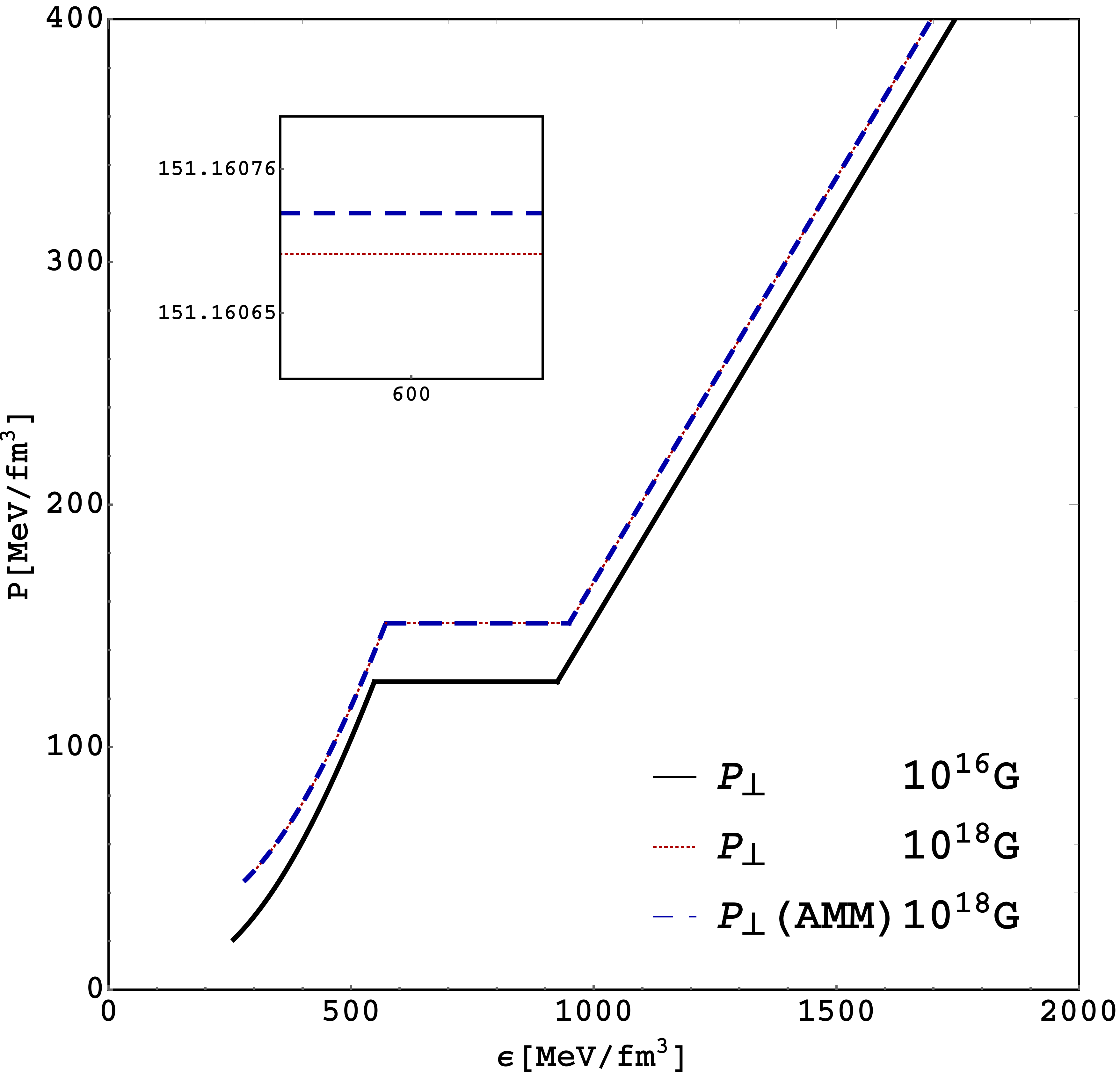}\\
(a) & (b)  \\
  \\
  \end{tabular}
    \end{center}
    \caption{(Color online) Parallel (a) and perpendicular (b) EOS in the Maxwell construction including (blue-dashed) the neutron AMM at $B^{HP}=10^{18}$G and excluding the neutron AMM at $B^{HP}=10^{18}$G (red-dotted) and at 
    $B^{HP}=10^{16}$G (black-solid).}
     \label{EOS}
\end{figure}

In Fig. \ref{EOS}, the parallel and perpendicular EOS are plotted while again considering the affects of the AMM at $10^{18}$G. As was the case with the normalized baryon density in Fig. \ref{Pressures}, the energy density experiences a jump across the phase-transition point, while the pressures remain continuous. This is indicative of the first order nature of the phase transition because the energy density harbors a first derivative of the free energy via the number density term found in (\ref{energy-density-1}). Even at a field strength of $10^{18}$G the AMM seems to play an insignificant role in the structure of the EOS in the Maxwell construction. This is due to the fact that at those field values the pure magnetic-field effect of the Maxwell pressure sweeps away the contribution to the pressure of the neutron AMM. 

\begin{figure}[t]
\begin{center}
\includegraphics[width=0.5\textwidth]{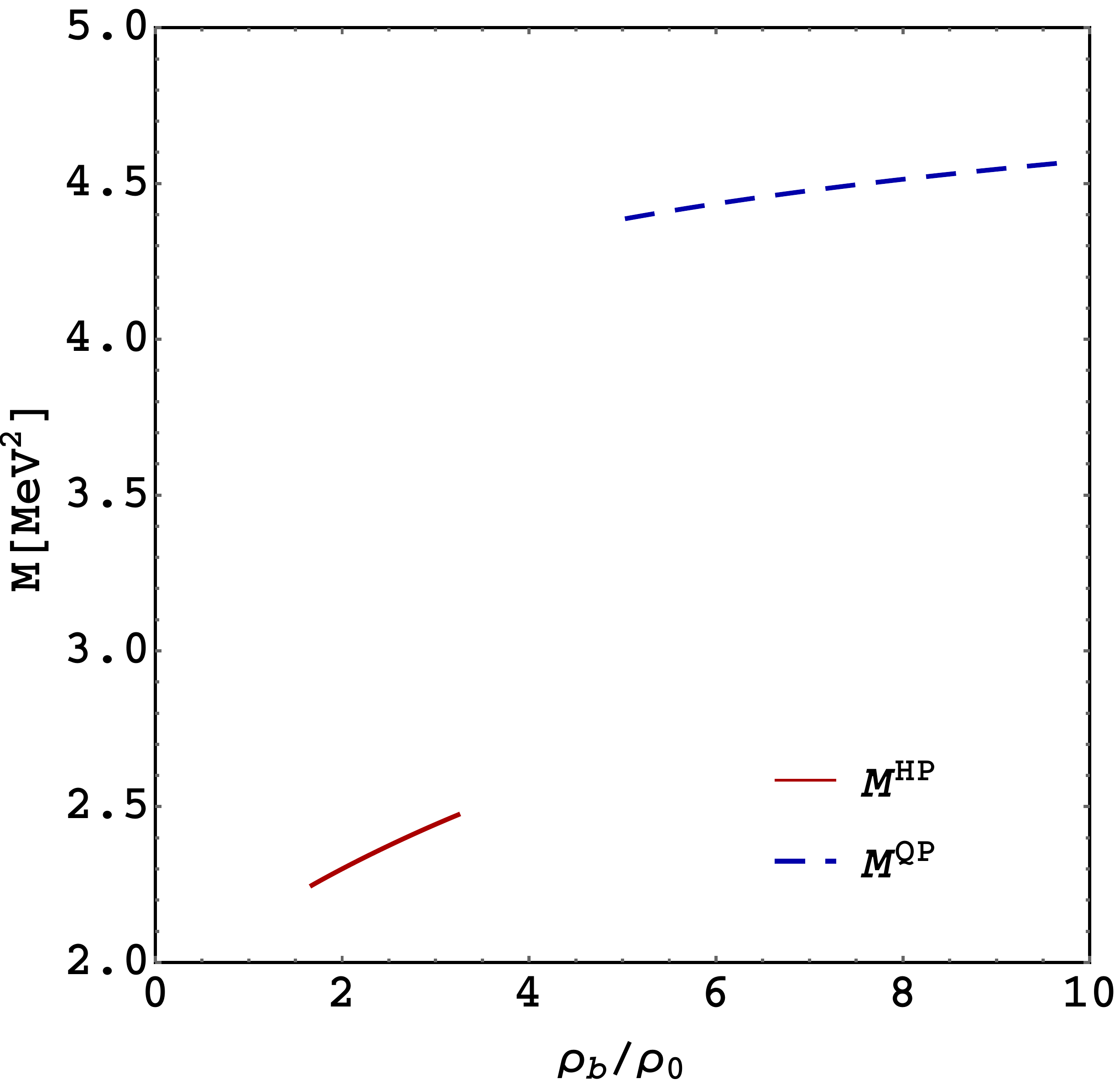}
\caption{(Color online) Magnetization, $M$, in the HP (red-solid) and QP (blue-dashed) versus baryonic charge density, $\rho_b$ normalized against the baryonic saturation density $\rho_0=0.153$ fm$^{-3}$, at $B^{HP}=10^{16}$ G.}
\label{MVrho}
\end{center}
\end{figure}

The system magnetization versus baryonic charge density is plotted in Fig. \ref{MVrho}  for $B^{HP}=10^{16}$ G. The solid line corresponds to the hadronic phase and the dashed line to the quark phase. It can be noticed that both parameters are discontinuous at the transition point as is expected for a first derivative of the thermodynamic potential in a first-order phase transition. It can be seen that the system magnetization increases when transitioning from the hadronic to the quark phase, which indicates that quark matter is more paramagnetic. The discontinuity observed for the magnetization around the phase transition in Fig. \ref{MVrho} confirms what we indicated in Section III-C regarding its implication for a jump in the magnetic field in transitioning from one phase to the other.

Hence, as was shown in Section III-C, the magnetic field after the phase transition to the quark phase should vary in strength with respect to the hadron phase. In our calculation we found that even for fields up to $10^{18}$ G the field variation is relatively small (i.e. $\Delta B_{max}=B^{QP}-B^{HP}\simeq 4.11\times10^{15}$ G) and consequently the magnetic monopole charge density that can accumulate at the phase transition boundary will also be relatively small for the systems under consideration. At first glance, it looks to be in contradiction with the fact that there is a significant jump in the magnetization (see Fig. 3). But, as we will see in the following section, the jump in $M$ is mainly due to the variation of the magnetic susceptibility during the transition (both of them, $M$ and $\chi$, double in value); while the strong-magnetic-field's main contribution to the phase transition equilibrium equations is driven by the Maxwell terms  (i.e. $B^2 \gg MB$) that do not differentiate the phase peculiarities (see Eqs. (\ref{Pressures-EOS-1}), (\ref{Par-Pres}) and (\ref{Per-Pres})). 

\begin{figure}[t]
\begin{center}
\includegraphics[width=0.5\textwidth]{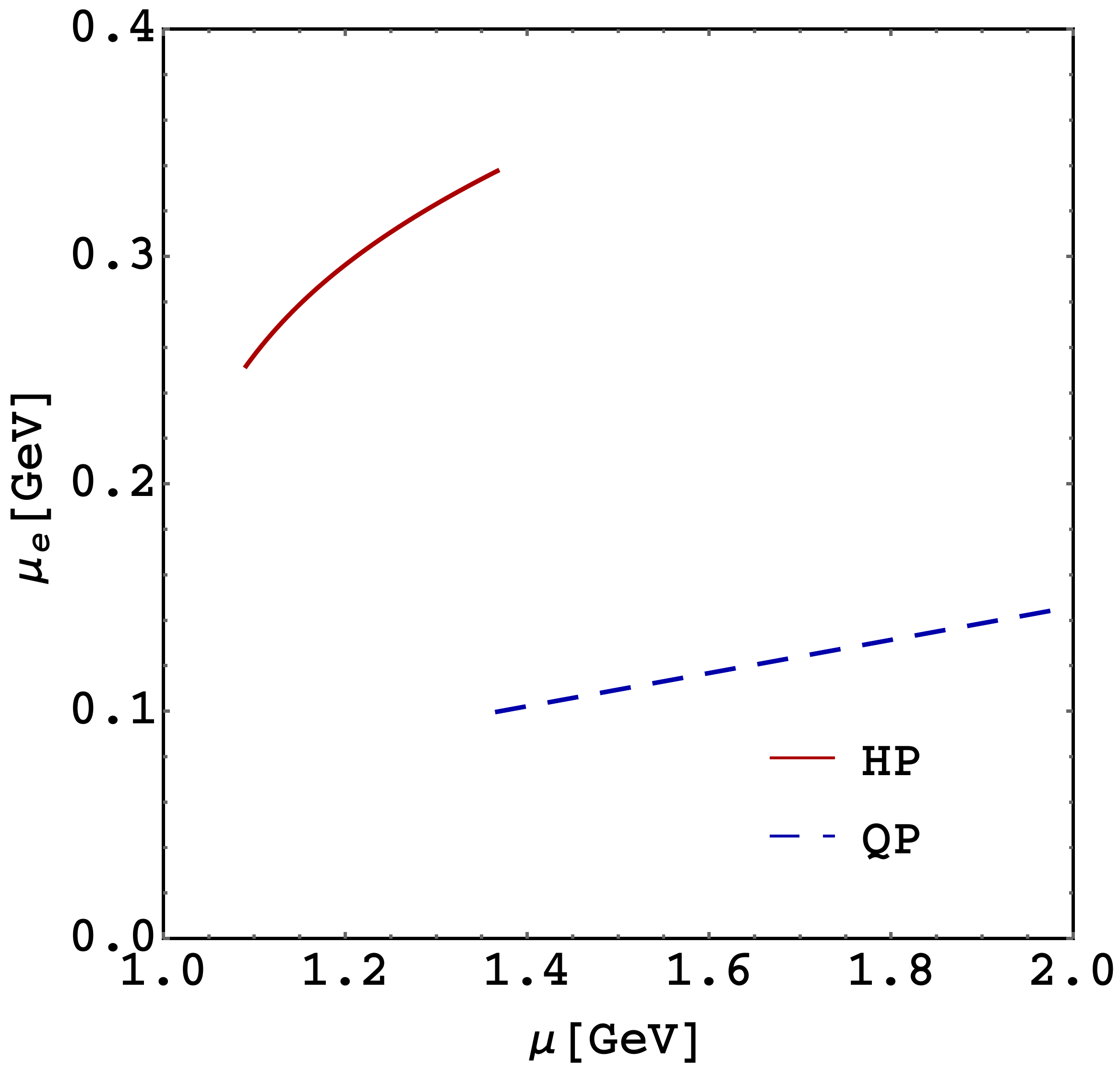}
\caption{(Color online) Electric chemical potential, $\mu_e$, in the HP (red-solid) and QP (blue-dashed) versus baryonic chemical potential, $\mu$, at $B^{HP}=10^{16}$G.}
\label{mueVmub}
\end{center}
\end{figure}

The electric chemical potential versus the baryonic chemical potential is represented by the solid line for the hadronic phase and by the dashed line for the quark phase in Fig. \ref{mueVmub}. The electric chemical potential jumps down when passing from the hadronic phase to the quark phase. The higher chemical potential in the hadronic phase indicates that a larger number of electrons is needed to compensate for the protons' electric charge than is needed to compensate for the quarks' electric charge. This discontinuous behavior for $\mu_e$ is typical of the Maxwell construction where the local neutrality conditions imply the existence of two independent chemical potentials for each phase. 


\section{Magnetic Susceptibility of the hadronic and Quark phases}\label{section4}

As follows from (\ref{B-Jump-3}),  the jump in the magnetic field, in going from one phase to the other, partially depends on the difference between the magnetizations of the two phases, or in other words, on which phase is more paramagnetic. With the objective of understanding the situation between the HP and the QP, we investigate in this section  the magnetic susceptibility of the two phases.

The magnetic susceptibility ($\chi_M$) of a system is defined as the coefficient of the linear expansion of the magnetization in powers of the magnetic field,
\begin{equation}\label{MSus}
	M=\chi_MB.
\end{equation}
The magnetization in the HP can be computed by differentiating the field dependent thermodynamic potentials 
$\Omega_p^{\mu}$ and $\Omega_e^{\mu}$ in (\ref{ThermoPotentialFiniteDensity-WFA}) with respect to $B$,
\begin{equation}\label{magnetizationHS}
	\begin{aligned}
&M_f^{\mu HP}=-\frac{\partial{{\Omega_f^{\mu}}}^{HP}}{\partial{B}}=\frac{e^2B}{6\pi^2}\ln\left[\frac{\left({\mu_p^*}+\sqrt{{\mu_p^*}^2-{m_p^*}^2}\right)\left({\mu^{HP}_e}+\sqrt{({\mu^{HP}_e})^2-{m_e}^2}\right)}{{m_p^*m_e}}\right],
\end{aligned}
\end{equation}
and the magnetic susceptibility is therefore 
\begin{equation}\label{MSusHS}
	\chi_M^{{\mu}HP}=\frac{e^2}{6\pi^2}\ln\left[\frac{\left({\mu_p^*}+\sqrt{{\mu_p^*}^2-{m_p^*}^2}\right)\left({\mu^{HP}_e}+\sqrt{({\mu^{HP}_e})^2-{m_e}^2}\right)}{{m_p^*m_e}}\right].
\end{equation}

\begin{figure}[t] \label{M-S}
\begin{center}
\includegraphics[width=0.6\textwidth]{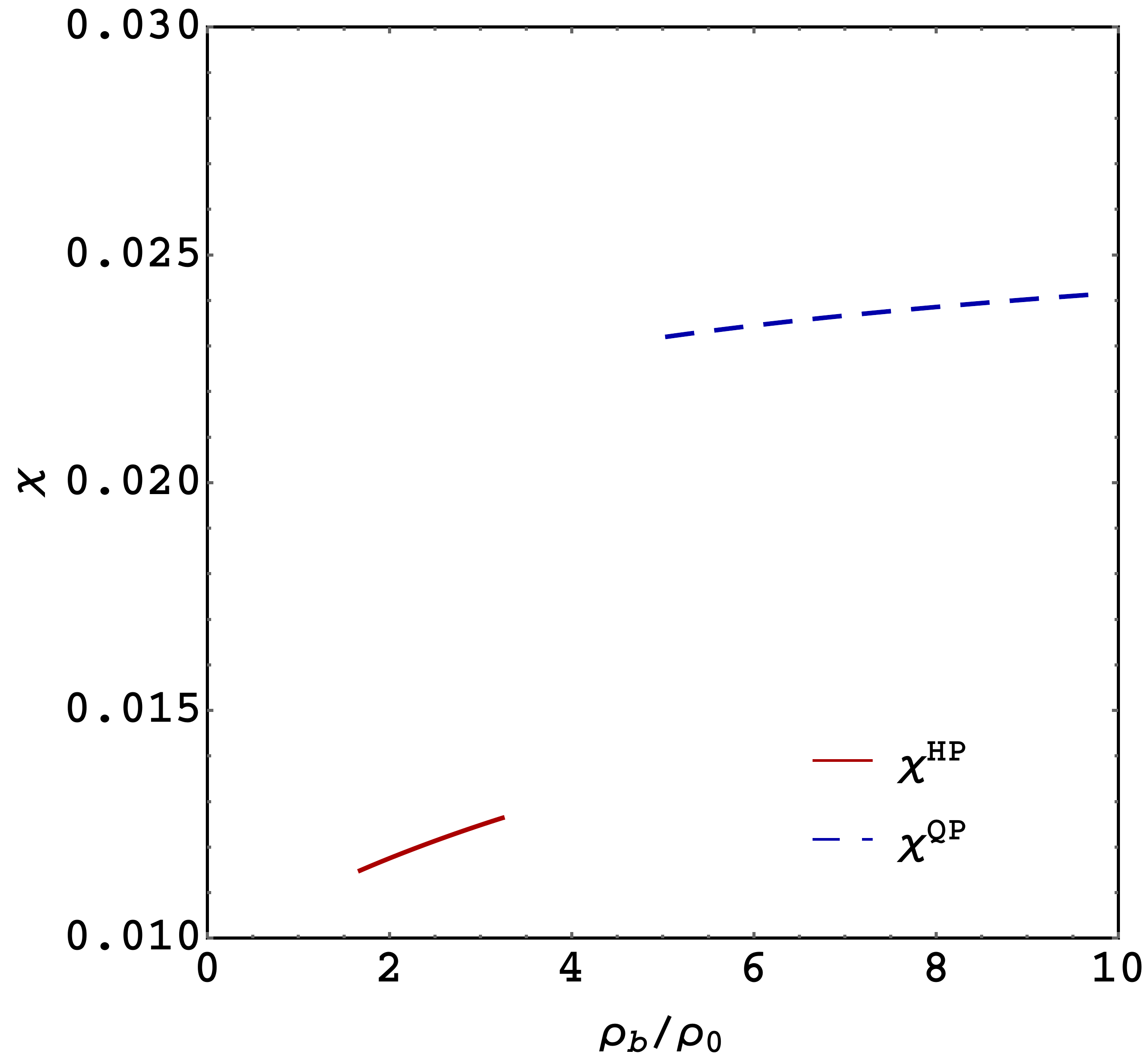}
\caption{(Color online) Magnetic susceptibility, $\chi$, in the HP (red-solid) and QP (blue-dashed) as a function of baryonic charge density, $\rho_b$ normalized against the baryonic saturation density $\rho_0=.153$ fm$^{-3}$, at $B^{HP}=10^{16}$ G.}
\label{chiVrho}
\end{center}
\end{figure}

Similarly, the QP magnetization is given by 
\begin{equation}\label{magnetizationQS}
	\begin{aligned}
&M_f^{{\mu}QP}=\frac{e^2B}{18\pi^2}\ln\left[\frac{\left({\mu_u}+\sqrt{{\mu_u}^2-{m_u}^2}\right)^4\left({\mu_d}+\sqrt{{\mu_d}^2-{m_d}^2}\right)\left({\mu^{QP}_e}+\sqrt{({\mu^{QP}_e})^2-{m_e}^2}\right)^{3}}{{m_u}^4{m_d}{m_e}^{3}}\right],
\end{aligned}
\end{equation}
and the corresponding magnetic susceptibility is
\begin{equation}\label{MSusQS}
	\begin{aligned}
&\chi_M^{{\mu}QP}=\frac{e^2}{18\pi^2}\ln\left[\frac{\left({\mu_u}+\sqrt{{\mu_u}^2-{m_u}^2}\right)^4\left({\mu_d}+\sqrt{{\mu_d}^2-{m_d}^2}\right)\left({\mu^{QP}_e}+\sqrt{({\mu^{QP}_e})^2-{m_e}^2}\right)^{3}}{{m_u}^4{m_d}{m_e}^{3}}\right].
\end{aligned}
\end{equation}

In Fig. 5, the magnetic susceptibility $\chi_M$ is plotted across the phase transition  against baryon chemical potential at a hadronic magnetic field value of $B^{HP}=10^{16}$ G. The positivity of $\chi$ in both sectors suggests that both media are paramagnetic with stronger paramagnetism occurring in the QP. This supports the finding that the QP withstands  moderately stronger magnetic fields.


\section{Concluding remarks}\label{section6}

In this paper we investigated the hadron-quark first-order phase transition, which may take place in the dense core of NS in the presence of a magnetic field. Considering that a uniform magnetic field produces anisotropic EOS, we proposed a new set of anisotropic equilibrium conditions, which will drive the phase transition at finite densities of the magnetized phases. To put forward this approach we 
extended the study of the anisotropic EOS in the presence of a uniform magnetic field that was initially investigated in Ref. \cite{magnetizedfermions, neutrons} for fermions, by including different species of mesons. 

We found that the thermodynamic pressures in the presence of a magnetic field coincide with those found from the components of the quantum-statistical average of the energy-momentum tensor, having a component along the magnetic field, $p_\|$, which is given by the Pauli pressure, while the component transverse to the magnetic field, $p_\bot=p_\|-MB$, has an extra term related to the energy generated by the interaction of the system magnetic moment (i.e. the system magnetization) and the magnetic field.

 Our results show that the magnetic field is boosted to achieve a moderately larger value in the quark phase. For example, in the simple model we are considering for free quarks, for a magnetic field of $10^{18}$ G in the hadronic phase, the field in the quark phase is increased by $4.11 \times 10^{15}$ G. This increase is due to an increase in the magnetic susceptibility after deconfinement, as is shown in Fig. \ref{chiVrho}.
 
 We also notice that in transitioning from the hadronic to the quark phase, by increasing the magnetic field in the hadronic phase the critical value of the chemical potential for the phase transition increases by a small amount. The increase is not significant because if the corresponding thermodynamic potentials of the systems under consideration are expanded in the Lorentz scalars depending on the magnetic field and the chemical potentials entering in the formulation, $\Omega \sim c_1 \mu^4+c_2\mu_e^4+c_3 B^2$, there are no products of $\mu$ with $B$. Then, if the difference between $B^{HP}$ and $B^{QP}$ is very small the Maxwell contributions cancel out in the mechanical equilibrium equations (\ref{Par-Pres}) and (\ref{Per-Pres}) and hence the solution for $\mu_c$ does not depend strongly on the magnetic field.
 Nevertheless, we expect that other more realistic quark phases that can be realized at moderate densities, such as the magnetic dual chiral density wave (MDCDW) phase \cite{Frolov} - \cite{MDCDW}, where the magnetization has an anomalous component that is linearly dependent on the baryonic chemical potential, can produce a more significant jump in the magnetic field as can be seen from (\ref{B-Jump-3}). We add that in this case the term mixing $\mu$ with $B$ comes from a chiral anomaly that allows for a term in the thermodynamic potential proportional to a Lorentz scalar depending on the totally antisymmetric Levi-Civita tensor $\epsilon_{\mu\nu\rho\lambda}$, which can be contracted with the medium four velocity $u_\mu$, the magnetic field given through $F^{\mu\nu}$ and the modulation vector $b_\mu$. 
 
 Another important result we are reporting is that as a consequence of the magnetic field discontinuity around the phase-transition boundary a density of magnetic monopoles with a net magnetic charge is created. The mechanism producing this effect is based on the dual Schwinger pair production and the magnetic charge gradient produced by the difference between the magnetic field strengths  of the two magnetic fields positioned on both sides of the phase-transition border. 
 
 We should mention that the approach we are introducing here to deal with the change of phase from hadronic to quark matter can be useful when studying the mass-radius relationship of hybrid stars. There already exists a discussion about this issue (see Ref. \cite{Stefano}, for example), but the calculations that have been done up to now, assumed that the magnetic field is much smaller than the baryonic chemical potential, which implies that the anisotropy in the EOS is not relevant. Nevertheless, for $B\gtrsim \mu^2$, the anisotropy becomes significant and should be considered in the formalism together with a modification of the TOV equations that would reflect the anisotropy in the space-time metric. That is, the TOV equations need to be derived for a metric that is in agreement with the axial symmetry of the problem instead of the spherical symmetry, which is assumed in the present form. This new required formulation of the TOV equations is an open non-trivial question that has to be solved.

Finally, we want to point out that the general results we are reporting here for the anisotropic hadron-quark first-order phase transition in the presence of a magnetic field will gain more interest when other dense quark matter phases, such as the MDCDW phase \cite{Frolov, MDCDW},  for instance, is taken into account.  In another direction, it will also be of interest to investigate the effects of gluons in this scenario. It has recently been found that the inclusion of gluons in the color superconducting quark phase modifies the EOS so that it becomes softer, thus opening the possibility of decreasing the maximum stellar mass that can be reached by strange stars \cite{CS-Gluons}. Moreover, it has been known for some time, that gluon condensates generated by the non-linear nature of the gluon interactions, as in particular the $A_4$=constant condensate, can play a role in QCD \cite{Gluon-Cond}. It will be then of interest to investigate what role, if any, they will have in the EOS and hence in the hadron-quark phase transition of dense QCD in the presence of a magnetic field.

\begin{appendix}


\section{Meson Contribution to the Stress Energy Tensor} \label{Appendix-A}

In \cite{magnetizedfermions} the EOS for a magnetized system of charged fermions was derived by using quantum-statistical methods. In this Appendix we extend those results to the case where mesons are also present as is the case in the hadron phase described by the NLW model.

Following the approach used in \cite{magnetizedfermions}, the components of the quantum-statistical average of the energy-momentum tensor are given by 
\begin{equation}\label{QSaverage}
	T^{\mu\nu}=\frac{1}{\beta{V}}\langle\hat{\tilde{\tau}}^{\mu\nu}\rangle=\frac{1}{\beta{V}}\frac{\Tr[\hat{\tilde{\tau}}^{\mu\nu}e^{-\beta(\hat{H}-\sum_i\mu_i \hat{N}_i)}]}{\mathcal{Z}}
\end{equation}
where
\begin{equation}\label{Hat-Tilde}
		\hat{\tilde{\tau}}^{\mu\nu}=\int_0^{\beta}d\tau\int{d^3x}\hat{\tau}^{\mu\nu}(\tau,x),
\end{equation}
$\hat{H}$ is the Hamiltontian, $\hat{N}$ is the particle number, $\mu$ is the chemical potential, $\hat{\tau}^{\mu\nu}$ are components of the field-theoretic energy-momentum tensor operator, and $\mathcal{Z}$ is the partition function of the grand canonical ensemble

\begin{equation}
	\mathcal{Z}=\Tr{e^{-\beta(\hat{H}-\sum_i\mu_i \hat{N}_i)}}.
\end{equation}

Given a theory incorporating fermion, meson, and Maxwell fields with corresponding Lagrangian density 
\begin{equation}
	\mathcal{L}=\mathcal{L}_f+\mathcal{L}_m+\mathcal{L}_{M}.
\end{equation}
(\ref{QSaverage}) may be expressed as a path integral with
\begin{equation}\label{QSaveragePI}
\begin{aligned}
	\beta{V}T^{\mu\nu}&=\left<\hat{\tilde{\tau}}^{\mu\nu}_f\right>+\left<\hat{\tilde{\tau}}^{\mu\nu}_{M}\right>+\left<\hat{\tilde{\tau}}^{\mu\nu}_{\phi_m}\right>
\\
&=\frac{\int\mathcal{D}A_{\mu}\mathcal{D}\phi_m\mathcal{D}\psi\mathcal{D}\bar{\psi}\left(\tilde{\tau}^{\mu\nu}_f+\tilde{\tau}^{\mu\nu}_{M}+\tilde{\tau}^{\mu\nu}_{m}\right)e^{-\int_0^{\beta}d\tau\int{d^3x}{\mathcal{L}_E(\tau,x)}}}{\mathcal{Z}},
\end{aligned}
\end{equation}
where $A_{\mu}$, $\phi_m$, and $\psi$ denote any number of Maxwell, meson, and fermion fields respectively. $\mathcal{L}_E$ is the Euclidean effective Lagrangian with imaginary time given by 
\begin{equation}
	\mathcal{L}_E=-\mathcal{L}(t\to-i\tau)-\sum_i \mu_i{\rho}_i,
\end{equation}
with $\rho_i$ being the conserved particles' number densities. The components $\tau^{\mu\nu}$ are determined by calculating the general relativistic matter stress energy tensor 
\begin{equation}\label{grstressenergy}
	\tau^{\mu\nu}=\frac{-2}{\sqrt{-g}}\frac{\delta}{\delta{g}_{\mu\nu}}\left(\sqrt{-g}\mathcal{L}\right)
\end{equation}
and afterwards reverting back to a local Minkowskian reference frame $g^{\mu\nu}\to\eta^{\mu\nu}$. We note here that $\mathcal{L}_{\psi}$ in general contains the fermion fields $\psi$ as well as the meson $\phi_m$ and Maxwell fields $A_\mu$. In \cite{magnetizedfermions} the first two terms of the rhs of (\ref{QSaveragePI}) were determined without any meson interaction, so it remains to calculate the contribution of the mesons entering in $\mathcal{L}_m$ and $\mathcal{L}_{\psi}$. 

In the path integral formalism the partition function is given by
\begin{equation}
	\mathcal{Z}=\int\mathcal{D}A_{\mu}\mathcal{D}\phi_m\mathcal{D}\psi\mathcal{D}\bar{\psi}e^{-\int_0^{\beta}d\tau\int{d^3x}\left({{\mathcal{L}_f}_E+{\mathcal{L}_m}_E}+{\mathcal{L}_M}_E\right)}.
\end{equation}

In the MFA the $A_{\mu}$ and $\phi_m$ fields are replaced by their average values and the corresponding functional integrals are removed. The partition function in the MFA then becomes
\begin{equation}
	\mathcal{Z}=\int\mathcal{D}\psi\mathcal{D}\bar{\psi}e^{-\int_0^{\beta}d\tau\int{d^3x}\left[{{\mathcal{L}_f}_E(\psi,\bar{\phi}_m, \bar{A}_{\mu})+{\mathcal{L}_m}_E}(\bar{\phi}_m)+{\mathcal{L}_M}_E(\bar{A}_{\mu})\right]},
\end{equation}

\begin{equation}
	=e^{{-\beta{V}{\mathcal{L}_m}_E}(\bar{\phi}_m)}e^{-\beta{V}{\mathcal{L}_M}_E(\bar{A}_{\mu})}\int\mathcal{D}\psi\mathcal{D}\bar{\psi}e^{-\int_0^{\beta}d\tau\int{d^3x}\left[{\mathcal{L}_f}_E(\psi,\bar{\phi}_m, \bar{A}_{\mu})\right]}.
\end{equation}
Where $\bar{\phi}_m$  denotes the meson expectation values to be determined by the minimum equations and $\bar{A}_\mu$ the external electromagnetic field.

The thermodynamic potential of the grand canonical ensemble is given by $\Omega=-\frac{1}{\beta{V}}\ln\mathcal{Z}$. Then in the MFA the thermodynamic potential can be decomposed as follows 
\begin{equation}\label{A-11}
	\Omega=-\mathcal{L}_m(\bar{\phi}_m)-\mathcal{L}_M(\bar{A}_{\mu})-\frac{1}{\beta{V}}\ln\int\mathcal{D}\psi\mathcal{D}\bar{\psi}e^{-\int_0^{\beta}d\tau\int{d^3x}\left[{\mathcal{L}_\psi}_E(\psi,\bar{\phi}_m, \bar{A}_{\mu})\right]},
\end{equation}

\begin{equation}
	=\Omega_m+\Omega_M+\Omega_{f},
\end{equation}
where $\Omega_m=-\mathcal{L}_m(\bar{\phi}_m)$ and $\Omega_M=-\mathcal{L}_M(\bar{A}_{\mu})$. Note that in \cite{magnetizedfermions} it was found that $\Omega_M=B^2/2$ for a uniform background magnetic field directed along the $z$ direction. The explicit form of $\Omega_m$ for the meson fields considered in this paper is given in the next section. As can be seen from (\ref{A-11}), the meson fields enter in the baryon Lagrangian $\mathcal{L}_b$ and the pure meson Lagrangian $\mathcal{L}_m$. We must show how each contributes to the stress energy tensor (\ref{grstressenergy}).

\subsection{Pure Meson Contribution}

Using the meson Lagrangian from (\ref{LagrangianHadrons-2}), acknowledging that we will eventually make the mean field approximation where derivatives of the fields vanish, and making the dependence on the metric explicit we get 
\begin{equation}
	\tilde{\mathcal{L}}_m=-\frac{1}{2}m_{\sigma}^2\sigma^2-U(\sigma)+\frac{1}{2}m_{\omega}^2\omega_{\lambda}\omega_{\tau}g^{\tau\lambda}+\frac{1}{2}m_{\rho}^2\rho_{\lambda}\rho_{\tau}g^{\tau\lambda}.
\end{equation}

Further using 
\begin{equation}
	\frac{\delta\sqrt{-g}}{\delta{g_{\mu\nu}}}=\frac{1}{2}\sqrt{-g}g^{\mu\nu},\:\:\frac{\delta{g^{\tau\lambda}}}{\delta{g^{\mu\nu}}}=-g^{\mu\tau}g^{\nu\lambda}
\end{equation}

we find that 
\begin{equation}
\tau_m^{\mu\nu}=\frac{-2}{\sqrt{-g}}\frac{\delta}{{\delta}g_{\mu\nu}}\left(\sqrt{-g}\tilde{\mathcal{L}}\right)
\end{equation} 

\begin{equation}
	\tau_{m}^{\mu\nu}=\left(\frac{1}{2}m_{\sigma}^2\sigma^2+U(\sigma)-\frac{1}{2}m_{\omega}^2\omega_{\lambda}\omega^{\lambda}-\frac{1}{2}m_{\rho}^2\rho_{\lambda}\rho^{\lambda}\right)\eta^{\mu\nu}+m_{\omega}^2\omega^{\mu}\omega^{\nu}+m_{\rho}^2\rho^{\mu}\rho^{\nu}.
\end{equation}

Keeping in mind that the only nonvanishing expectation values of the vector mesons are their zeroth components we have that 
\begin{equation}
\begin{aligned}
T^{00}_m&=\frac{1}{2}m_{\sigma}^2\bar{\sigma}^2+U(\bar{\sigma})+\frac{1}{2}m_{\omega}^2\bar{\omega}_0^2+\frac{1}{2}m_{\rho}^2\bar{\rho}_{0}^2,
\\
T^{ii}_m&=-\frac{1}{2}m_{\sigma}^2\bar{\sigma}^2-U(\bar{\sigma})+\frac{1}{2}m_{\omega}^2\bar{\omega}_{0}^2+\frac{1}{2}m_{\rho}^2\bar{\rho}_{0}^{2}.
\end{aligned}
\end{equation}

This can be expressed in a fully covariant form as 
\begin{equation}\label{puremesonset}
	T^{\mu\nu}_m=\Omega_m\eta^{\mu\nu}+\left(\bar{\omega}_0{\rho}_{\omega}+\bar{\rho}_0{\rho}_{\rho}\right)u^{\mu}u^{\nu},
\end{equation}
where for the $\sigma$, $\omega_{\mu}$, and $\rho_{\mu}$ meson fields in the MFA,
\begin{equation}\label{Omega-m}
	\Omega_m=\frac{1}{2}m_{\sigma}^2\bar{\sigma}^2+U(\bar{\sigma})-\frac{1}{2}m_{\omega}^2\bar{\omega}_{0}^2-\frac{1}{2}m_{\rho}^2\bar{\rho}_{0}^{2},
\end{equation}
and ${\rho}_{\omega}=-(\partial\Omega_m/\partial{\bar{\omega}_0})$ and ${\rho}_{\rho}=-(\partial\Omega_m/\partial{\bar{\rho}_0})$.

\subsection{Baryon-Meson Contribution}

In the case of the fermion fields, spinors are introduced into curved spacetime by considering an orthonormal tetrad basis $\varepsilon_{m}$ in the tangent space $\textbf{T}_p$ of each point $p$ of the spacetime manifold $\mathcal{M}$. The tetrad basis is related to the coordinate basis $e_{\mu}$ through the Vierbein $V^m\,_{\mu}$ matrix by
\begin{equation}
	e_{\mu}=V^{m}\,_{\mu}\varepsilon_m,\quad\quad{g_{\mu\nu}}=\eta_{mn}V^m\,_{\mu}V^n\,_{\nu}.
\end{equation}

Here greek indices are lowered and raised by acting with $g_{\mu\nu}$, while latin indices are lowered and raised by acting with the flat spacetime metric $\eta_{mn}$ and its inverse. $\mathcal{L}_b$ is expressed in curved spacetime as 
\begin{equation}\label{lagrangiancurvedspacetime}
	\mathcal{L}_b^{c.s.}=\frac{i}{2}\left(\bar{\psi_b}\gamma^{\mu}\nabla_{\mu}\psi_b-\nabla_{\mu}\bar{\psi}_b\gamma^{\mu}\psi_b+2im^*_b\bar{\psi}_b\psi_b+2i\bar{\psi}_b\gamma^{\mu}{X_b}_{\mu}\psi_b\right)
\end{equation}
where $\nabla_{\mu}$ is the covariant derivative in curved spacetime, $\gamma^{\mu}=e_m\,^{\mu}\gamma^m$ are generalizations of the Dirac matrices $\gamma^m$ in curved spacetime, and $X_{b\mu}$ is given by
\begin{equation}
	{X_b}_{\mu}=q_b\bar{A}_{\mu}+g_{\omega{b}}\bar{\omega}_{\mu}+g_{\rho{b}}{\tau_{3}}_b \bar{\rho}_{\mu}.
\end{equation} 
Eq. (\ref{grstressenergy}) can be equivalently expressed by taking the variation with respect to the Vierbein fields

\begin{equation}
	\tau^{\mu\nu}_b=-\frac{V^{m\mu}}{V}\frac{\delta\left(V\mathcal{L}^{c.s.}_b\right)}{\delta{V^m\,_{\nu}}}
\end{equation}
where $V=\det\left(V^n\,_{\alpha}\right)$. Carrying out the variation of (\ref{lagrangiancurvedspacetime}) with respect to $V^m\,_{\nu}$ we arrive at 
\begin{equation}
\begin{aligned}
	{\tau_b}_{\mu\nu}=&-ig_{\mu\nu}\left[\bar{\psi}_b\gamma^{\lambda}\nabla_{\lambda}\psi_b-\nabla_{\lambda}\bar{\psi}_b\gamma^{\lambda}\psi_b\right]+\frac{i}{2}\left[\bar{\psi}_b\gamma_{(\mu}\nabla_{\nu)}\psi_b-\nabla_{(\mu}\bar{\psi}_b\gamma_{\nu)}\psi_b\right]
\\
&+{g}_{\mu\nu}\bar{\psi}_b\left[m_b^*+\gamma^{\lambda}{X_b}_{\lambda}\right]\psi_b-\bar{\psi}_b\gamma_{\nu}{X_b}_{\mu}\psi_b.
\end{aligned}
\end{equation}

Then, to get the Euclidean expression for $\tau^{\mu\nu}_b$ in the grand canonical ensemble we take $g_{\mu\nu}\to\eta_{\mu\nu}$, $t=-i\tau$, and ${X_b}^{0}\to{{X}_b}^{0}-\mu_b$, which yields
\begin{equation}\label{fermionset}
\begin{aligned}
	\tau^{\mu\nu}_b(\tau,x)=&-i\eta^{\mu\nu}\bar{\psi}_b\left(i\gamma^{0}\partial_{\tau}+\gamma^i\partial_i\right)\psi_b+\frac{i}{2}\bar{\psi}_b\left(\gamma^{\mu}d^{\nu}+\gamma^{\nu}d^{\mu}\right)\psi_b
\\
&+{\eta}^{\mu\nu}\bar{\psi}_b\left[m_b^*+\gamma^{\lambda}{X_b}_{\lambda}\right]\psi_b-\bar{\psi}_b\gamma^{\nu}{X_b}^{\mu}\psi_b.
\end{aligned}
\end{equation}

Here, integration by parts was used to transfer the derivatives from $\bar{\psi}_b$ to $\psi_b$ and $d_{\mu}=(i\partial_{\tau},\partial_i)$. Working in the Landau gauge $A_{\mu}=Bx_1\delta_{\mu2}$ and taking into account that in the MFA the only nonvanishing components of the vector mesons are the zero components, and the only nonvanishing components of $X_b^{\mu}$ are $X_b^0$ and $X_b^2$, (\ref{fermionset}) becomes
\begin{equation}
\begin{aligned}
	\tau^{00}_b&={\mathcal{L}_b}_E-\bar{\psi}_b\gamma^0\partial_{\tau}\psi_b+\bar{\psi}_b\mu_b^*\gamma^0\psi_b
\\
	\tau^{11}_b&=-{\mathcal{L}_b}_E+i\bar{\psi}_b\gamma^1\partial^1\psi_b
\\
	\tau^{22}_b&=-{\mathcal{L}_b}_E+i\bar{\psi}_b\gamma^2\partial^2\psi_b-\bar{\psi}_b\gamma^2q_bBx^1\psi_b
\\
	\tau^{33}_b&=-{\mathcal{L}_b}_E+i\bar{\psi}_b\gamma^3\partial^3\psi_b,
\end{aligned}
\end{equation}
where ${\mathcal{L}_b}_E$ is the baryon contribution to $\mathcal{L}_E$. $\tau^{ii}_b$ have the same form as was found in \cite{magnetizedfermions}, but $\tau^{00}_b$ has a new structure entering as a shift in the chemical potential $\mu_b\to\mu_b^*$. So, we focus on $\tau^{00}$ only. Making the variable change $s=\tau/\beta$ we have that
\begin{equation}
	\beta\frac{\partial{\mathcal{Z}_b}}{\partial{\beta}}=\int\mathcal{D}\psi_b\mathcal{D}\bar{\psi}_b\left[-\beta\int_0^{1}ds\int{d^3}x{\mathcal{L}_b}_E+\beta\int_0^{1}ds\int{d^3}x\bar{\psi}_b\frac{1}{\beta}\gamma^0\partial_{s}\psi_b\right]e^{-\beta\int_0^{1}ds\int{d^3}x{\mathcal{L}_b}_E},
\end{equation}

\begin{equation}
	=\int\mathcal{D}\psi_b\mathcal{D}\bar{\psi}_b\left[\int_0^{\beta}d\tau\int{d^3}x\left(-{\mathcal{L}_b}_E+\bar{\psi}_b\gamma^0\partial_{\tau}\psi_b\right)\right]e^{-\int_0^{\beta}d\tau\int{d^3}x{\mathcal{L}_b}_E},
\end{equation}

\begin{equation}
	=\int\mathcal{D}\psi_b\mathcal{D}\bar{\psi}_b\left[-\tilde{\tau}^{00}_b+\int_0^{\beta}d\tau\int{d^3}x\bar{\psi}_b\left(\mu-g_{\omega{b}}\bar{\omega}^0-g_{\rho{b}}{\tau_3}_b\bar{\rho}^0\right)\gamma^0\psi_b\right]e^{-\int_0^{\beta}d\tau\int{d^3}x{\mathcal{L}_b}_E}.
\end{equation}

Hence,
\begin{equation}\label{t00average}
	\langle\hat{\tilde{\tau}}^{00}_b\rangle=-\frac{\beta}{\mathcal{Z}_b}\frac{\partial{\mathcal{Z}_b}}{\partial{\beta}}+\beta\mu_b^*\langle\hat{N}_b\rangle.
\end{equation}
where $N_b=\int{d^3}x\bar{\psi}_b\gamma^0\psi_b$. Using the grand canonical potential $\Phi_b=-\frac{1}{\beta}\ln\mathcal{Z}_b$ and the fact that $\langle{\hat{N}_b}\rangle=-(\partial\Phi_b/\partial\mu_b)_{T,V}$, (\ref{t00average}) can be expressed as 
\begin{equation}
	\langle\hat{\tilde{\tau}}^{00}_b\rangle=-\frac{\partial\Phi_b}{\partial{T}}+\beta\Phi_b-\beta\mu_b^*\frac{\partial\Phi_b}{\partial\mu_b}.	
\end{equation}

Then using the fact that the thermodynamic potential is given by $\Omega_b=\Phi_b/V$, we arrive at the baryon contribution to the energy density.
\begin{equation}
	\varepsilon_b=\frac{1}{\beta{V}}\langle\hat{\tilde{\tau}}^{00}\rangle=\Omega_b+TS_b+\mu^*_b{\rho}_b,
\end{equation}
where $S_b=-(\partial\Omega_b/\partial{T})_{V,\mu}$and ${\rho}_b=-(\partial\Omega_b/\partial{\mu})_{V,T}$ are baryon entropy and baryon number density corresponding to species $b$ respectively. Following \cite{magnetizedfermions} we can write the baryon quantum statistical average of the stress energy tensor in the following covariant form

\begin{equation}\label{baryonset}
	T^{\mu\nu}_b=\frac{1}{\beta{V}}\langle\hat{\tilde{\tau}}^{\mu\nu}_b\rangle=\Omega_b\eta^{\mu\nu}+\left(\mu^*_b{\rho}_b+TS_b\right)u^{\mu}u^{\nu}+BM_b\eta_{\perp}^{\mu\nu}.
\end{equation}


\section{Charged Fermion Thermodynamic Potential in the Weak-Field Approximation}\label{Appendix-B}

The finite density thermodynamic potential for charged fermions without B-AMM interaction takes the form 
\begin{equation}\label{ChargedFermionTP-1}
\Omega^{\mu} =-\frac{eB}{(2\pi)^2}\int_{-\infty}^{\infty}dp_3\left\{\left(\mu-E_0\right)\Theta\left(\mu-E_0\right)+2\displaystyle\sum_{l=1}^{\infty}\left(\mu-E\right)\Theta\left(\mu-E\right)\right\},
\end{equation}
where the notation $E_0=E(l=0)$ has been used. This can be rewritten as 
\begin{equation}\label{ChargedFermionTP-2}
\Omega^{\mu} =-\frac{2eB}{(2\pi)^2}\int_{-\infty}^{\infty}dp_3\displaystyle\sum_{l=0}^{\infty}\left(\mu-E\right)\Theta\left(\mu-E\right)+\frac{eB}{(2\pi)^2}\int_{-\infty}^{\infty}dp_3\left(\mu-E_0\right)\Theta\left(\mu-E_0\right).
\end{equation}

If we consider $E$ to be a function of $u=2eBl$, then for small $B$ the separation between consecutive values of $u$ is also small. This allows us to use the Euler-Maclaurin formula \cite{EulerMaclaurin} without remainder 
\begin{equation}\label{EulerMaclaurinA}
	\displaystyle\sum_{l=0}^{\infty}f\left(eBl\right)\approx\frac{1}{eB}\int_0^{\infty}f(x)dx+\frac{f(\infty)+f(0)}{2}+\displaystyle\sum_{k=1}^{\infty}\frac{B_{2k}}{(2k)!}(eB)^{2k-1}\left[f^{(2k-1)}\left(\infty\right)-f^{(2k-1)}\left(0\right)\right].
\end{equation}
where $x$ is a continuous variable and $B_{2k}$ are the Bernoulli numbers to approximate the sum over Landau levels. Using (\ref{EulerMaclaurinA}) to approximate the sum in (\ref{ChargedFermionTP-2}) and keeping terms up to $\mathcal{O}\left(\left(eB\right)^2\right)$ we get 
\begin{equation}\label{ChargedFermionTP-3}
	\Omega^{\mu}\approx-2\int_{-\infty}^{\infty}\frac{d^3p}{(2\pi)^3}\left(\mu-E_N\right)\Theta\left(\mu-E_N\right)-\frac{(eB)^2}{3(2\pi)^2}\ln\left[\frac{\mu+\sqrt{\mu^2-m^2}}{m}\right].
\end{equation}

The first term comes from the continuous integral in (\ref{EulerMaclaurinA}) after making the sequence of substitutions: $u=2x$, $v^2=u$, and $p_1=v\cos\theta$, $p_2=v\sin\theta$. Here $E_N=\sqrt{p_1^2+p_2^2+p_3^2+m^2}$ is the energy spectrum for neutral free relativistic fermions and so the first term in (\ref{ChargedFermionTP-3}) is equivalent to the neutron contribution in (\ref{ThermoPotentialFiniteDensity-WFA}) with $\mu_n^*\to\mu$ and $m_n^*\to{m}$. Putting this all together (\ref{ChargedFermionTP-3}) becomes
\begin{equation}
	\Omega^{\mu}\approx\frac{-1}{24\pi^2}\left\{\left(2{\mu}^4-5{m}^2{\mu}^2\right)\sqrt{1-\left(\frac{m}{\mu}\right)^2}+\left(3{m}^4+2(eB)^2\right)\ln\left[\frac{{\mu}+\sqrt{{\mu}^2-{m}^2}}{{m}}\right]\right\}.
\end{equation}

\end{appendix}

\acknowledgments
This work was supported in part by NSF grant PHY-2013222.


\begin{thebibliography}{99}

\bibitem{Deconfinement} H. Satz, Nucl. Phys. A 418 (1984) 447.

\bibitem{Shapiro}S. L. Shapiro and S. A. Teukolsky, Black Holes, White Duarfs, and Neutron Stars (Wiley,New York, 1983).

\bibitem{Collins}J. C. Collins and M. J. Perry, Phys. Rev. lett. 30 (1975) 1353.

\bibitem{Soft-gamma  repeaters} Ibrahimet et al., Astrophys. J. 609 (2004) L21;  S. Kulkarni and D. Frail, Nature 365 (1993) 33; T. Murakamiet et al., Nature 368 (1994) 127; D. Marsden and J. C. Higdon, The Astro. Phys. Jour., 550 (2001) 397; R. Dib, V. M. Kaspi, and F. P. Gavriil,  The Astro. Phys. Jour.,  673  (2008) 1044.

\bibitem{Nuclear-B} M. Bocquet, S. Bonazzola, E. Gourgoulhon, and J. Novak, Astron. Astrophys. 301 (1995) 757; L. Dong and S. L. Shapiro, ApJ. 383 (1991) 745.

\bibitem{magnetizedfermions} E. J. Ferrer, V. de la Incera, J. P. Keith, I. Portillo, and P. L. Springsteen, Phys. Rev. C 82 (2010) 065802.

\bibitem{Magnetohydro-Eq} C. Y. Cardall, M. Prakashand J. M. Lattimer, Astrophys.J. 554 (2001) 322.
 
 \bibitem{Oertel} D. Chatterjee, T. Elghozi, J. Novac and M. Oertel, Mon. Not. Roy. Astron. Soc. 447 (2015) 3785.

 \bibitem{neutrons} E. J. Ferrer and A. Hackebill, Phys. Rev. C 99 (2019) 065803.  
 
\bibitem{First-Trans} S. D. H. Hsu and M. Schwetz, Phys. Lett. B 432 (1998) 2003; O. Scavenius, A. Mocsy, I. N. Mishustin and D. H. Rischke, Phys. Rev. C 64 (2001) 045202.

\bibitem{Agasian} N. O. Agasian and S. M. Fedorov, Phys. Lett. B, Vol. 663 (2008) 445.

\bibitem{Trans-B} D. Bandyopadhyay, S. Chakrabarty and S. Pal, Phys. Rev. Lett. 79 (1997) 2176;  M. Orsaria, I. F. Ranea-Sandoval, H. Vucetich, F. Weber, Int. J. Mod. Phys. E 20 (2011) Supp. 02, 25; R. H. Casali, L. B. Castro, D. P. Menezes, Phys. Rev. C 89 (2014) 015805; B. Franzon, V. Dexheimer and S. Schramm, Mon. Not. Roy. Astron. Soc. 456 (2016) 2937.

\bibitem{Rabhi} A. Rabhi, H. Pais, P. K. Panda and C. Providencia, J. Phys. G: Nucl. Paet. Phys. 36 (2009) 115204. 

\bibitem{Canuto}  V. Canuto and H. Y. Chiu, Phys. Rev. 173 (1968) 1210.

\bibitem{Glendenning} N. K. Glendenning, \textit{Compact Stars}, (2000) New York: Springer.

\bibitem{Walecka} B. D. Serot and J. D. Walecka, Adv. Nucl. Phys. 16 (1986) 1.

\bibitem{Frolov} I. E. Frolov, V. Ch. Zhukovsky and K. G. Klimenko, Phys. Rev. D 82 (2010) 076002.

\bibitem{Stefano} S. Carignano, E. J. Ferrer,  V. de la Incera, and L. Paulucci, Phys. Rev. D 92 (2015) 105018.

\bibitem{Inhomogeneous-phases} E. J. Ferrer, V. de la Incera and A. Sanchez, Acta Phys. Polon. Supp. 5 (2012) 679; M. Buballa and S. Carignano, Eur. Phys. J. A 52 (2016) 57.

\bibitem{MDCDW} T. Tatsumi, K. Nishiyama and S. Karasawa, Phys. Lett. B 743 (2015) 66;  E. J. Ferrer and V. de la Incera, Phys. Lett. B 769 (2017) 208; Nucl. Phys. B931 (2018) 192; Universe 4 (2018) 54; B. Feng, E. J. Ferrer, I. Portillo, Phys. Rev. D   \textbf{101} (2020) 056012;  E. J. Ferrer and V. de la Incera, Universe 7 (2021) 458.

\bibitem{MCFL} L. Paulucci, E. J. Ferrer, V. de la Incera, and J. E.  Horvath, Phys. Rev. D  83 (2011) 043009; E. J. Ferrer, V. de la Incera and C. Manuel, J. Phys. A 39 (2006) 6349; E. J. Ferrer and V. de la Incera, Lect. Notes Phys. 871 (2013) 399; arXiv:1208.5179 [nucl-th].

\bibitem{London} F. London, Superfluids, John Wiley and Sons, New York, (1950). L. Onsager, Proceeding of the International Conference on Theoretical Physics, 935, Science Council of Japan, Tokyo, (1954).

\bibitem{M-Flux-Spin} M. Saglam and B. Boyacioglu, Intern. J. of Mod. Phys. B,16 (2002) 607; M. Saglam and B. Boyacioglu, Phys. Stat. Sol. B 230 (2002) 133.

\bibitem{Magnetic-Flux} A. Holas, S. Olszewski and  D. Pftrsch, Z. Naturforsch. 45 a (1990) 847; O. Yilmaz, M. Saglam and Z. Z. Aydin, Concepts of Physics 4 (2007) 141.

\bibitem{Magnetic-Flux-Experiment} B. S. Deaver and W. M. Fairbank, Phys. Rev. Lett. 7 (1961) 43; R. Doll and M. Nabauer, ibid, 7 (1961) 51.

\bibitem{First-order} S. Klevansky, Rev. Mod. Phys. 64 (1992) 649;
G. F. Burgio, M. Baldo, P. K. Sahu, and H.-J. Schulze, Phys. Rev. C 66 (2002) 025802;
I. N. Mishustin, M. Hanauske, A. Bhattacharyya, L. M. Satarov, H. Stocker, and W. Greiner, Phys. Lett. B 552  (2003) 1;
P. Costa, M. C. Ruivo and C. A. de Sousa, Phys. Rev. D77 (2008) 096001;
J. D. Carroll, D. B. Leinweber, A. G. Williams, and A. W. Thomas, Phys. Rev. C 79 (2009) 045810;
P. Costa, M. C. Ruivo, C. A. de Sousa and H. Hansen, Symmetry 2 (2010) 1338;
S. Weissenborn, I. Sagert, G. Pagliara, M. Hempel, and J. Schaffner-Bielich,  Astrophys. J. 740 (2011) L14;
T. Fischer,et al., Astrophys. J. Suppl.194  (2011) 39;
H.-J. Schulze and T. Rijken, Phys. Rev. C 84 (2011) 035801;
C. A. Graeff, M. D. Alloy, K. D. Marquez, C. Provid$\hat{e}$ncia, D. P. Menezes, JCAP 01 (2019) 024;
D. Sen and T. K. Jha, J. Phys. G 46 (2019) 015202.

\bibitem{Greiner}A. Bhattacharyya, I. N. Mishustin and W. Greiner, J. Phys. G 37 (2010) 025201.

\bibitem{Gibbs Cons} N. K. Glendenning, Phys. Rev. D 46 (1992) 1274.

\bibitem{MC} S. Schramm, V. Dexheimer, and R. Negreiros, Eur. Phys.J. A 52 (2016) 14.
 D. Logoteta, and I. Bombaci, Phys. Rev. D 88 (2013) 063001;
B. Franzon, R. O. Gomes and S. Schramm, Mon. Not. Roy. Astron. Soc. 463 (2016) 571.

\bibitem{Surface-Tension}T. Norsen and S. Reddy, Phys. Rev. C 63 (2001) 065804; D. N. Voskresensky, M. Yasuhira and T. Tatsumi, Phys. Lett. B 541 (2002) 93.

\bibitem{MC-Arguments} T. Maruyama, T. Tatsumi, T. Endo and S. Chiba, Recent Res. Devel. Phys. 7 (2006) 1.

\bibitem{Sigma-Estimates} E. Farhi and R. L. Jaffe, Phys. Rev. D 30 (1984) 2379; M. S. Berger and R. L. Jaffe, Phys. Rev C 35 (1987) 213; K. Kajantie, L. Karkainen and K. Rummukainen, Nucl. Phys. B 357 (1991) 693; S. Huang, J. Potvion, C. Rebbi and S. Sanielevici, Phys. Rev. D 423 (1991) 2056.

\bibitem{Lugones} A. G. Grunfeld and G. Lugones, Astron. Nachr. 342 (2021) 205.

\bibitem{Tatsumi} T. Maruyama, S. Chiba, H-J. Schulze and T. Tatsumi, Phys. Rev. D 76 (2007) 123015.

\bibitem{Fukushima} H-L Chen, K. Kukushima, X-G Huang and K. Mameda, Phys. Rev. D 96 (2017) 054032.

\bibitem{Manton} I. K. Affleck and N. S. Manton, Nucl.  Phys. B 194 (1982) 38; I. K. Affleck, O. Alvarez,  and N. S. Manton, Nucl. Phys.B 197 (1982) 509. 

\bibitem{Schwinger}J. S. Schwinger, Phys. Rev. 82 (1951)  664.

\bibitem{Gould}O. Gould and A. Rajantie, Phys. Rev. Lett. 119 (2017) 241601; Phys. Rev. D 96  (2017) 076002

\bibitem{MP-NS}     P. V. S. Pavan Chandra, M. Korwar, A. M. Thalapillil,  Phys. Rev. D 101 (2020)  075028.

 \bibitem{MM-Accumulation}  A. Rajantie, Phil. Trans. Roy. Soc. Lond. A 377 (2019) 2161, 20190333. Contribution to: Topological avatars of new physics.

\bibitem{Table} N. K. Glendenning and S. A. Moszkowski, Phys. Rev. Lett. 67 (1991) 2414.

\bibitem{insignificance} E. J. Ferrer, V. de la Incera, D. Manreza Paret,  A. Perez Martinez, and A. Sanchez, Phys. Rev. D 91 (2015) 085041.

\bibitem{Broderick} A. Broderick, M. Prakash and J. M. Lattimer, The Astrophys.l Jour. 537 (2000) 351.

\bibitem {CS-Gluons} E. J. Ferrer, V. de la Incera and L. Paulucci, Phys. Rev. D 92 (2015) 043010.

\bibitem{Gluon-Cond} M. A. Shifman et al.  Nucl. Phys. B 147 (1979) 385;
G. K. Savvidy, Phys. Lett. B71 (1977) 133;
N. K. Nielsen and P. Olesen, Nucl. Phys. B 144 (1978) 376;
H. B. Nie!sen and M. Ninoxriya, Nucl. Phys. B 153 (1980) 57; ibid. Bl69 (1980) 309; J.  Ambjom and P. Olesen, Nucl. Phys. B 170 (1980) 265;
A. Cabo and A. E. Shabad, Trud. FIAN SSSR 111 (1979); A. Cabo, S. Penaranda and R. Martinez, Mod. Phys. Lett. A10 (1995) 2413;
M. Ninomiya and N. Sakai, Nucl. Phys. B 190 (1981) 316;
R. Anishetty, J. Phys. G. Nucl. Phys. 10 (1984) 423; 
K. J. Dahlern, Z. Phys. C 29 (1988) 553;
V. M. Belyaev and V. L. Eletsky, Z. Phys. C 45 (1990) 355;
K. Enqvist and K. Kajantie, Z. Phys. C 47 (1990) 291;
V. M. Belyaevm, Phys. Lett. B 241 (1990) 91;
V. V. Skalozub and I. V. Chub, Yad. Fiz. 57 (1994) 344; O. A. Borisenko, J. Bohacik and V. V. Skalozub, Fortschr. Phys. 43 (1995) 01;
D. Ebert, V. Ch. Zhukovsky, and  A. S. Vshivtsev, Inter. J. of Mod. Phys. A 13 (1998) 1723. 

\bibitem{EulerMaclaurin} G. Arfken, \textit{Mathematical Methods for Physicists}, 3rd ed. (Academic, New York, 1985).


 \end{thebibliography}
\end{document}